\documentclass[12pt,preprint]{aastex}

\usepackage{natbib}\usepackage{graphicx}

\newcommand\bb[1] {   \mbox{\boldmath{$#1$}}  }
\newcommand\del{\bb{\nabla}}
\newcommand\bcdot{\bb{\cdot}}
\newcommand\btimes{\bb{\times}}
\newcommand{\gs}{\raisebox{-.8ex}{$\buildrel{\textstyle>}\over\sim$}}

\shorttitle{3D evolution of self--gravitating magnetized disks}
\shortauthors{Fromang et al.}

\begin{document}

\title{
Evolution of self--gravitating magnetized disks. II- Interaction between 
MHD turbulence and gravitational instabilities}

\author{
S\'ebastien Fromang\altaffilmark{1},
Steven A.\ Balbus\altaffilmark{2,3}, 
Caroline Terquem\altaffilmark{1,4} 
and Jean--Pierre De Villiers\altaffilmark{2}}

\altaffiltext{1}{Institut d'Astrophysique de Paris, 98Bis Bd Arago,
75014
Paris, France }
\altaffiltext{2}{Virginia Institute of Theoretical Astronomy, Department
of Astronomy, University of Virginia, Charlottesville, VA 22903--0818,
USA}
\altaffiltext{3}{Laboratoire de Radioastronomie, Ecole Normale 
Sup\'erieure, 24 rue Lhomond, 75231 Paris Cedex 05, France}
\altaffiltext{4}{Universit\'e Denis Diderot--Paris VII, 2 Place
Jussieu, 75251 Paris Cedex 5, France}

\begin{abstract}
We present 3D magnetohydrodynamic (MHD) numerical simulations of the
evolution of self--gravitating and weakly magnetized disks with an
adiabatic equation of state.  Such disks are subject to the development
of both the magnetorotational and gravitational instabilities,
which transport angular momentum outward.  As in previous studies,
our hydrodynamical simulations show the growth of strong $m=2$ spiral
structure.   This spiral disturbance drives matter toward the central
object and disappears when the Toomre parameter $Q$ has increased
well above unity.  When a weak magnetic field is present as well, the
magnetorotational instability grows and leads to turbulence.  In that
case, the strength of the gravitational stress tensor is lowered by
a factor of about~2 compared to the hydrodynamical run and oscillates
periodically, reaching very small values at its minimum.  We attribute
this behavior to the presence of a second spiral mode with higher pattern
speed than the one which dominates in the hydrodynamical simulations.
It is apparently excited by the high frequency motions associated with
MHD turbulence.  The nonlinear coupling between these two spiral modes
gives rise to a stress tensor that oscillates with a frequency which is
a combination of the frequencies of each of the modes.  This interaction
between MHD turbulence and gravitational instabilities therefore results
in a smaller mass accretion rate onto the central object.  \end{abstract}

\keywords{accretion, accretion disks - MHD - self--gravity}

\section{Introduction}

In systems such as the disks surrounding low mass protostars
or active galactic nuclei, the simultaneous appearance of both
gravitational and magnetic instabilities is expected.  During the
first stages of their evolution, for example, protoplanetary disks
are expected to be rather massive because of strong infall from the
parent molecular cloud.  As the disk builds up in mass as a result
of the collapse of an envelope, its surface mass density becomes
large enough for gravitational instabilities to develop (e.g.,
\citeauthor{laughlin&bodenheimer94}~\citeyear{laughlin&bodenheimer94}).
These disks are also believed to be sufficiently ionized, at
least over some extended regions, to be coupled to a magnetic field
\citep{gammie96,sanoetal00,fromang02}.

By modeling the outer parts of disks around quasi-stellar objects
(QSOs) as steady, viscous, geometrically thin, and optically thick,
\citet{goodmanj03} has argued that they are self-gravitating.  More
precisely, he predicts self--gravitational instabilities to develop
beyond about $10^{-2}$~parsecs from the central object.  In addition,
it has been suggested by \citet{menou&quatert01} that self-gravitating
regions of disks around QSOs are likely to be coupled to a magnetic field.

The stability of a thin, self-gravitating gas disk is controlled by the 
Toomre $Q$ parameter \citep{toomre64}:

\begin{equation}
Q=\frac{c_s\kappa}{\pi G \Sigma} \, ,
\end{equation}

\noindent
where $c_s$ is the sound speed, $\kappa$ is the epicyclic frequency
(see, e.g.,
\citeauthor{binney&tremaine87}~\citeyear{binney&tremaine87}), 
$\Sigma$ is the disk surface mass
density and $G$ is the gravitational constant.  Gaseous disks are
unstable against axisymmetric perturbations when $Q \le 1$, and
against non-axisymmetric perturbations when $Q \gs 1$.  

Since analytical predictions of the nonlinear evolution of
gravitational instabilities are difficult, there have been a large
number of numerical simulations of gravitationally unstable
disks. Despite the rather daunting technical problems of combining
three-dimensional (3D) hydrodynamic calculations with rapid and
accurate Poisson equation solvers, significant progress have been
made. To do so, the energetics must be treated crudely, with the focus
squarely on purely dynamical behavior. Using this strategy, the above
$Q$ criterion for instability has been confirmed (and shown to
still be approximately valid for disks of finite thickness), and the
properties of the unstable modes have been studied as a function
of the disk parameters \citep{tohline&hachisu90,woodward94}.
Several authors have investigated the saturation properties of
the instability, and have shown that it is capable of transporting
significant amount of mass and angular momentum in a few orbital times
\citep{pap&savonije91,laughlin&bodenheimer94,pickett96,laughlin97}.
The first calculations mostly used simple adiabatic equations of
state (EOS).  More recently, isothermal disks have also been studied
\citep{pickett98,pickett00a,boss98,mayer02}.  Some new investigations
also include a simplified treatment of the disk radiative cooling
\citep{pickett03,rice03,boss02}.

All these models were purely hydrodynamical, and neglected the effect of
magnetic fields.  However, it is known that stability of astrophysical
disks is extremely sensitive to the presence of weak magnetic fields.
In particular, the magnetorotational instability (MRI) completely disrupts
laminar Keplerian flow when a subthermal magnetic field of any geometry
is present. This was first understood by \citet{balbus&hawley91}. Since
then, it has been shown through many numerical simulations that the
nonlinear outcome of the MRI is MHD turbulence, which, in common with
gravitational instabilities, transports angular momentum outward (see
\citeauthor{balbus&hawley98} \citeyear{balbus&hawley98}, or \citeauthor{balbusaraa03} \citeyear{balbusaraa03},  for a review).  Since disks
around low--mass stars and around QSOs may be both magnetized and
self--gravitating, the spiral structure gravitational
transport described above must somehow develop in a medium in the throes of
MHD turbulence.

The question naturally arises as to how these two powerful instabilities
interact with one another.  What is the ultimate effect on the global
properties of accretion disks, and in particular, on the critical
transport properties of mass and angular momentum?  To keep this
initial investigation tractable, we must restrict ourselves here to
an adiabatic EOS.  But the dynamical behavior of ``simple'' adiabatic
disks is still rich, and contains unanticipated findings. In a companion
paper to this one (\citeauthor{fromangetal04a}~\citeyear{fromangetal04a},
hereafter paper~I), we carried out 2D axisymmetric numerical simulations
of the evolution of massive and magnetized disks. The results show that
the MRI behaves in a self--gravitating environment as it does in zero
mass disks. Turbulent transport of angular momentum causes the disk
to evolve toward a two component structure: (1) an inner thin disk in
Keplerian rotation fed by (2) an outer thick disk whose rotation profile
deviates from Keplerian, strongly influenced by self-gravity.  However,
angular momentum transport by gravitational instabilities cannot develop
in axisymmetric simulations, which leaves unanswered the question of
the outcome of the interaction between both instabilities. This is the
subject of the present paper.

The plan of the paper is as follows: in section 2, we present our
numerical methods. The initial state of our simulations will be described
in section 3. We present our results in section 4 and, finally, give
our conclusions in section 5.

\section{Numerical methods}

\subsection{Algorithms}

The calculations in this paper are based on the
equations of ideal MHD:

\begin{eqnarray}
\frac{\partial \rho}{\partial t} + \del \bcdot (\rho {\bf v})  =  0, \\
\rho \left( \frac{\partial {\bf v}}{\partial t} + {\bf v} \bcdot \del
{\bf v} \right)  =  - \del P - \rho \del \Phi + \frac{1}{4
\pi} (\del \btimes {\bf B}) \btimes {\bf B}, \\
\rho \left( \frac{\partial }{\partial t} + {\bf v} \bcdot \del \right)
\left( \frac{e}{\rho} \right)  =  -P \del \bcdot {\bf v}, \\
\frac{\partial {\bf B}}{\partial t}  =  \del \btimes ( {\bf v} \btimes
{\bf B} ),
\label{MHD equations}
\end{eqnarray}

\noindent
where $\rho$ is the mass density, $e$ is the energy density, $\bf{v}$
is the fluid velocity, $\bf{B}$ is the magnetic field, $P$ is the gas
pressure and $\Phi=\Phi_s+\Phi_c$ is the total gravitational potential,
which has contributions $\Phi_s$ from the disk self--gravity and
$\Phi_c$ from a central mass. The Poisson equation determines the
gravitational
potential,
\begin{equation}
\nabla^2 \Phi_s = 4 \pi G \rho,
\end{equation}
and to close our system of equations, we
adopt an adiabatic equation of state for a monoatomic gas:

\begin{equation}
P = (\gamma -1)e, \quad \gamma = 5/3 .
\label{EOS}
\end{equation}

\noindent
To solve these equations, we use the GLOBAL code \citep{hawley&stone95}.
This uses standard cylindrical coordinates $(r, \phi, z)$ and
time--explicit Eulerian finite differences.  The magnetic field is
evolved using the combined Method of Characteristics and Constrained
Transport algorithm (MOC--CT), which preserves the divergence of the
magnetic field to machine accuracy.
Finally, we use outflow boundary
conditions in the radial and vertical directions, and periodic
boundary conditions in $\phi$.

In its original form, GLOBAL did not include a Poisson solver, and the
development of such a routine represents a major technical component of
the results we report here. 
The calculation is done in two steps. The potential $\Phi_s$ is first
computed at the grid boundary, using the spectral decomposition decribed
below, and then calculated on the whole grid using a very rapid method.
It is the first step, the boundary calculation,
that is computationally expensive.

In the expansion of $\Phi_s$, we have adopted the method of
\citet{cohl&tohline99}, which uses
half--integer Legendre functions in the Green's function.
This method is better suited to
cylindrical coordinates than the traditional expansion in spherical
harmonics, which are of course tailored to spherical coordinates.
Following \citet{cohl&tohline99}, $\Phi_s$ may be written

\begin{equation}
\Phi_s (r,\phi,z) = - \frac{G}{\pi \sqrt{r}} \int_{V} d\tau' \;
\frac{\rho(r',\phi',z')}{\sqrt{r'}} \sum_{m=0}^{\infty} \epsilon_m
Q_{m-1/2}(\chi) \cos m(\phi - \phi') \; .
\label{gravpot calc}
\end{equation}

\noindent
Here, $d\tau' = r' dr' d\phi' dz'$ is the elementary volume element,
and the integral
is taken over the whole computational domain. $Q_{m-1/2}$
denotes the half--integer order Legendre function of the second type
\citep{mathfunc}.  The argument $\chi$ is a function of position:

\begin{equation}
\chi = \frac{r^2+r'^{2}+(z-z')^2}{2rr'} \, .
\end{equation}

\noindent
The Legendre functions are computed once at the beginning of each
simulation and stored in memory.  At each time step, we calculate
$\Phi_s$ using equation~(\ref{gravpot calc}), in which the sum over
$m$ is truncated at some upper value $m_{max}$.  We then calculate
$\Phi_{s}$ everywhere on the grid, using a combination of a Fourier
transform in $\phi$ and the 2D Successive Over Relaxation (SOR)
Method \citep{hirsh88} in the $(r,z)$ plane.  Although this is an
efficient method, the calculation of the self--gravitating potential
is still very demanding of computational resources.  For the resolution
$(N_r,N_{\phi},N_z)=(128,64,128)$ used in this paper, the time required
by the Poisson solver still represents $\sim 40\%$ of the computation
time for $m_{max}=8$.

\subsection{Diagnostics}

We introduce and define some key quantities that have been used to analyze
the results of the simulations.  We denote the ratio of the volume
averaged thermal pressure to the volume averaged magnetic pressure as
$\langle \beta \rangle$:

\begin{equation}
\langle \beta \rangle = \frac{\langle P \rangle}{\langle B^2/8\pi
\rangle} \, .
\end{equation}

\noindent
This parameter is used primarily as a measure of the initial magnetic
field strength.

In 3D numerical simulations of magnetized self--gravitating disks,
angular momentum is transported by the sum of the Maxwell, Reynolds,
and gravitational stress tensors.  Following \citet{balbus&pap99} and
\citet{hawley00}, we define the height and azimuthal averages (noted
with an overbar) of each these respective stresses as:

\begin{eqnarray}
T^{Max}_{r\phi}(r,t) & = & - \frac{\overline{B_rB_{\phi}}}{4\pi} \, ,\\
T^{Ren}_{r\phi}(r,t) & = & \overline{\rho v_r
v_{\phi}}-\frac{\overline{\rho v_r} \textrm{ } \overline{\rho
v_{\phi}}}{\overline{\rho}} \, ,\\
T_{r\phi}^{grav}(r,t) & = & \frac{1}{4\pi G}  \overline{
\frac{\partial \Phi_s}{\partial r} \frac{1}{r} \frac{\partial
\Phi_s}{\partial \theta} } \, .
\end{eqnarray}

\noindent
As in paper I, volume averages of these quantities will be denoted as
$\langle T^{Max}_{r\phi}\rangle(t)$, etc. Note that $T_{r\phi}^{grav}$
is 
associated with the gravitational
torque resulting from non-axisymmetric disk structure.
This quantity clear vanishes in an axisymmetric simulation
($m_{max}=0$).
In this case, the standard $\alpha$ parameter \citep{shakura&sunyaev73} can be
defined as the sum of the Maxwell and Reynolds stress tensors
normalized by the gas pressure:

\begin{equation}
\alpha(r,t)=\frac{T^{Max}_{r\phi}(r,t)+
T^{Ren}_{r\phi}(r,t)}{\overline{P(r,t)}}
\, .
\label{eq:alpha}
\end{equation}

\section{Initial model}

We start our simulations with a disk model which is as close as
possible to hydrostatic equilibrium:

\begin{equation}
- \del P - \rho \del \left( \Phi_s + \Phi_c \right) + \rho r \Omega ^2
  {\bb e_r} = {\bf 0} \, .
\label{hydrostatic equilibrium}
\end{equation}

\noindent
Here $\Omega$ is the angular velocity and $\bb{e_r}$ is the unit vector
in the radial direction.  The coordinate system has its origin on the
disk center.  The potential $\Phi_c$ is due to a central mass $M_c$.
We chose $M_c= 2 M_d$, where $M_d$ is the disk mass.
The initial disk model is gravitationally unstable.

Because of the presence of the disk self--gravity,
equation~(\ref{hydrostatic equilibrium}) has to be solved iteratively.
We use the Self--Consistent Field (SCF) iterative method developed by
\citet{hachisu86}.  In this method, the radial profile
of the angular velocity $\Omega$ or, equivalently, the specific
angular momentum $j=r^2 \Omega$, is specified.
Following \citet{pickett96}, we fix
$j(r)$:

\begin{equation}
j=j_{r0} \left(\frac{M_r+M_c}{M_d+M_c}\right)^{q}
\end{equation}

\noindent
where $j_{r0}$ and $q$ are constant, and $M_r$ is the disk mass within
radius $r$.  Setting $q=2$ gives a $j$ profile close to that used by
\citet{pickett96}.  We begin the iteration with an arbitrary mass
density
$\rho$, from which we can calculate $\Phi_s$.  From $\rho$ and the
above expression for $j$ we also calculate $\Omega$ (note that it
still depends on the constant $j_{r0}$). The
relation~(\ref{hydrostatic equilibrium}) is then integrated to give
the value of the enthalpy $h$:

\begin{equation}
h =\frac{5}{2}K \rho^{2/3} = 
C - \left( \Phi_s + \Phi_c \right) + \int r \Omega ^2 dr \, ,
\end{equation}

\noindent
where the constant $C$ and $j_{r0}$ are determined from the boundary
conditions $\rho=0$ at $\left( r=R_{in}, z=0 \right)$ and at $\left(
r=R_{out},z=0 \right)$. Here $R_{in}$ and $R_{out}$ are the radial
boundaries of the disk.  The new density field is then calculated from
$h$
using the normalizing condition that $\rho_{max}=1$, which determines
the polytropic constant $K$. Upon iterating this procedure, we
converge to a model very close to equilibrium.

\begin{figure}
\begin{center}
\epsscale{.4}
\plotone{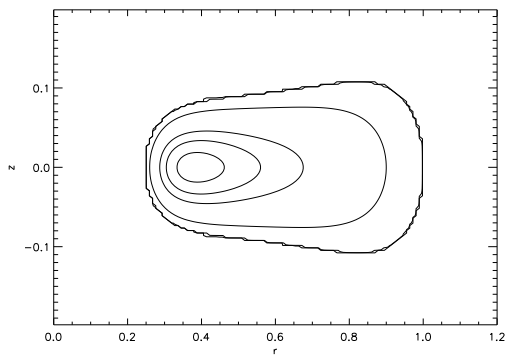}
\caption{Density contours in the $(r,z)$--plane of the initial disk
model used in all the simulations.  The contours shown are
$\rho=10^{-7}, 10^{-4}, 10^{-3}, 0.01, 0.1, 0.5, 0.7$ and $0.9$.  The
disk radial boundaries are $R_{in}=0.25$ and $R_{out}=1$.  The central
mass is twice that of the disk.}
\label{initial model}
\end{center}
\end{figure}

The resulting disk model (with $R_{in}=0.25$ and $R_{out}=1$) has an
$\Omega$ profile close to Keplerian and a density profile displayed in
figure~\ref{initial model}.  Note that the disk is rather thin, with
an aspect ratio $H/r$ varying between $0.1$ and $0.2$.

As noted above, the ratio $M_c/M_d$ is chosen in such a way that the
Toomre $Q$ parameter is initially close to unity.  The radial profile
of $Q$ in the initial disk model is shown in figure~\ref{initial
toomre}.  The minimum value of $Q$ is approximately $1.1$, and $Q$ is close to
unity over a large range of radii.  We therefore expect strong
non--axisymmetric gravitational instabilities to develop in this disk.

\begin{figure}
\begin{center}
\epsscale{.45}
\plotone{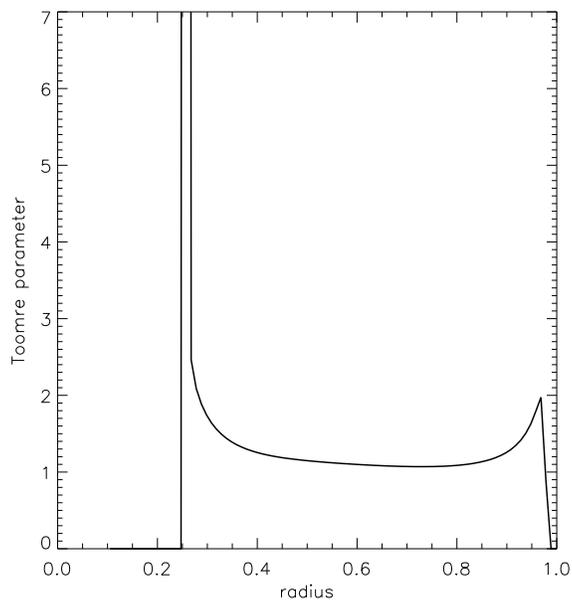}
\caption{Radial profile of the Toomre $Q$ parameter in the initial
disk model.  $Q$ is close to unity over a large range of radii.}
\label{initial toomre}
\end{center}
\end{figure}

\section{Results}

Table~\ref{models} lists the parameters of the different runs we
present below.  Column~$1$ gives the label of the model. HD refers to
a hydrodynamical run.  Models T and P start with a purely toroidal and
poloidal magnetic field, respectively.  Column~2 gives the
computational azimuthal domain and column~3 gives the highest fourier
component of the gravitational potential.  When $m_{max}=0$, i.e. when
only the $m=0$ component in the Fourier expansion of $\Phi_s$ is
included, gravitational instabilities cannot develop (recall that
$Q>1$, so that the disk is stable against axisymmetric perturbations).
Therefore, models T1 and P1 enable us to study the evolution of MHD
turbulence and to compare it with previous work and with the 2D
simulations of paper~I.  In models T2, T2$_{low}$, T2$^{*}$, T3 and P2, both
gravitational and magnetic instabilities develop.  In model T2$^{*}$,
the non--axisymmetric part of $\Phi_s$ is included only after $6$
orbits, i.e. after MHD turbulence has established itself.  Column~4
gives the ratio of the volume-averaged thermal and magnetic pressures
and column~5 gives the resolution $(N_r,N_{\phi},N_z)$ of the run.

\begin{table}
\begin{center}
\begin{tabular}{@{}llcccc}
\hline\hline
Model & $\phi$--range & $m_{max}$ & $\langle \beta \rangle$ & Resolution 
\\
\hline\hline
HD & $[0,\pi]$ & $8$ & $\infty$ & $(128,64,64)$ \\
\hline
T1 & $[0,\pi /2]$ & $0$ & $8$ & $(128,32,128)$ \\
T2 & $[0,\pi]$ & $8$ & $8$ & $(128,64,128)$ \\
T2$_{low}$ & $[0,\pi]$ & $8$ & $8$ & $(64,64,64)$ \\
T2$^*$ & $[0,\pi]$ & $0$--$8^{\dagger}$ & $8$ & $(128,64,128)$ \\
T3 & $[0,\pi]$ & $16$ & $8$ & $(128,64,128)$ \\
\hline
P1 & $[0,\pi /2]$ & $0$ & $300$ & $(128,32,128)$ \\
P2 & $[0,\pi]$ & $8$ & $300$ & $(128,64,128)$ \\
\hline\hline
\end{tabular}

\vspace{+0.1cm}
~~~ {\footnotesize $^\dagger$ For this run, $m_{max}=0$ when 
$t \in [0,5.8]$, while $m_{max}=8$ for $t>5.8$.}
\end{center}

\caption{Model parameters. Column~2 gives the computational azimuthal
domain, column~3 gives the highest Fourier component of the
gravitational potential included in the calculation, column~4 gives
the ratio of the volume averaged thermal and magnetic pressures and
column~5 gives the resolution $(N_r,N_{\phi},N_z)$ of the run.  Model
HD is hydrodynamical.  Models T and P start with a purely toroidal and
poloidal magnetic field, respectively.  When $m_{max}=0$, the disk
self--gravitating potential is forced to stay axisymmetric. In model
T2$^*$, $m_{max}=0$ at the beginning of the run and is set to 8 after
a few orbits.  }
\label{models}
\end{table}

In all the models, an adiabatic equation of state is used.  The
computational domain extends radially from 0.1 to 1.4, and vertically
from $-0.2$ to $0.2$.  In the azimuthal direction, the computational
domain extends from 0 to either $\pi/2$ or $\pi$.  The smaller range
is used in the $m_{max}=0$ gravitationally stable cases.  Indeed,
\citet{hawley00} and \citet{pap&nelson03a} have shown that an azimuthal
domain of $\pi/3$ is generally sufficient to describe the transport
properties of MHD turbulence.  When we allow for the development of
gravitational instabilities, we restrict the azimuthal domain to the half
disk $[0,\pi]$.  This saves computational time, but of course allows only
even modes to develop.  The focus of the paper is not on the detailed
spectrum of modes which appear in a given disk model, however, but on the
interaction between MHD turbulence and the largest scale gravitational
modes.  This interaction should not be particularly sensitive to whether
an integer number of modes exactly fits in the half disk.

Time is measured in units of the orbital period at the initial outer
edge $R_{out}=1$ of the disk model.  Typical simulations are carried
out for $8$ to $10$ orbits at this position.  This corresponds to
60--80 orbits at the initial disk inner edge.  The simulations are
seeded by adding to the mass density at $r> 0.4$ random perturbations
with a relative amplitude of $5 \times 10^{-3}$.

We now describe in turn the hydrodynamical run, the simulations with
only MHD turbulence, and the runs with both gravitational and magnetic
instabilities.

\subsection{Control Hydrodynamical Run: Model HD}

\begin{figure}
\begin{center}
\epsscale{.4}
\plotone{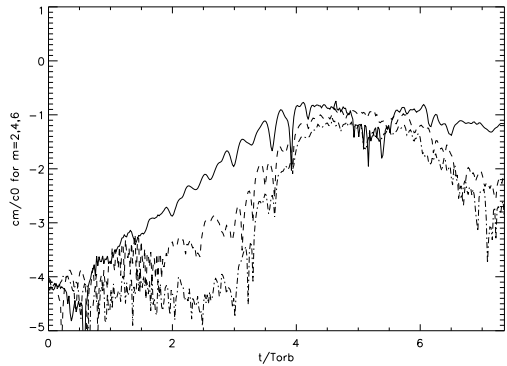}
\caption{Time evolution of the Fourier components of the density in
the equatorial plane for model HD.  The $y$--axis represents the
ratio of the amplitude of the $m$--th Fourier component of the
perturbed density to the unperturbed density.  From top
to bottom, the different curves correspond to the modes $m=2$ ({\it
solid line}), $m=4$ ({\it dashed line}) and $m=6$ ({\it dotted-dashed
line}). }
\label{time_fourier_hydro}
\end{center}
\end{figure}

The time evolution of the Fourier components of the density in the
equatorial plane is shown in figure~\ref{time_fourier_hydro} for the
modes $m=2, 4$ and~$6$ (from top to bottom).  The $m=2$
mode grows at the beginning of the simulation and saturates after $4$
orbits. Higher $m$ modes emerge after about $3$ orbits.  Apart from the
$m=4$ mode, which may also be linearly unstable, the $m>2$ modes appear
to be non--linearly excited.

The development of a $m=2$ spiral structure may be seen in
figure~\ref{image220 hydro}, which shows the logarithm of the density
in the equatorial plane at $t=4.27$.  (Note that the result of the
simulation has been extended by symmetry to cover the range $[0,2\pi]$).
This mode is clearly global.  Its pattern speed is $\Omega_p=6.28$, which
means that corotation (the radius where the gas angular velocity matches
the pattern speed) is located at the initial outer edge of the disk.
Such a mode is predicted to emerge by linear stability analyses of
self-gravitating disks \citep{pap&savonije91}.  The instability is due
to the interaction between waves that propagate near the outer boundary
and waves that reside inside the inner Lindblad resonance (where the
pattern speed in the frame corotating with the planet matches the gas
epicyclic frequency).  This is located at $r\sim 0.6$ in our disk model.

\begin{figure}
\begin{center}
\epsscale{.4}
\vspace{6cm}
\caption{Logarithm of the density in the equatorial plane for model~HD
at $t=4.27$. The linearly unstable two--arm global mode has become
non--linear.  It drives angular momentum outward and matter toward the
central point mass.}
\label{image220 hydro}
\end{center}
\end{figure}

During the simulation, matter is driven toward the disk center by the
gravitational torque associated with the spiral arms and, at $t \simeq
8$, $Q$ has become sufficiently high ($\gs 2$) that the disk
settles into a stable state.  The results of this simulation are in
agreement with theoretical expectations and with previous work, and
show that the Poisson solver performs satisfactorily in the hydrodynamical
regime.

\subsection{MHD Simulations in an Axisymmetric Gravitational Potential}

In the presence of a weak magnetic field, we expect our disk model to
be unstable to the MRI, regardless of the field geometry.
We first perform simulations in which only the MRI
develops (models T1 and P1).  This allows the
properties of the ensuing MHD turbulence to be quantified
and compared with previous work.
To prevent the growth of non--axisymmetric gravitational instabilities,
we retain only the $m=0$ component in the Fourier expansion of
$\Phi_s$.  The resolution is $(N_r,N_{\phi},N_z)=(128,32,128)$ and the
azimuthal domain extends from 0 to $\pi/2$, which would be equivalent to
a
resolution of $128^3$ over a range of $2\pi$.  

\subsubsection{Initial toroidal field: Model T1}

We add to the equilibrium disk model described above a  $\langle \beta
\rangle=8$ toroidal magnetic field  and run the simulation for about
$10$ orbits (about $80$ orbits at the initial disk inner edge).

\begin{figure*}
\begin{center}
\epsscale{.8}
\plotone{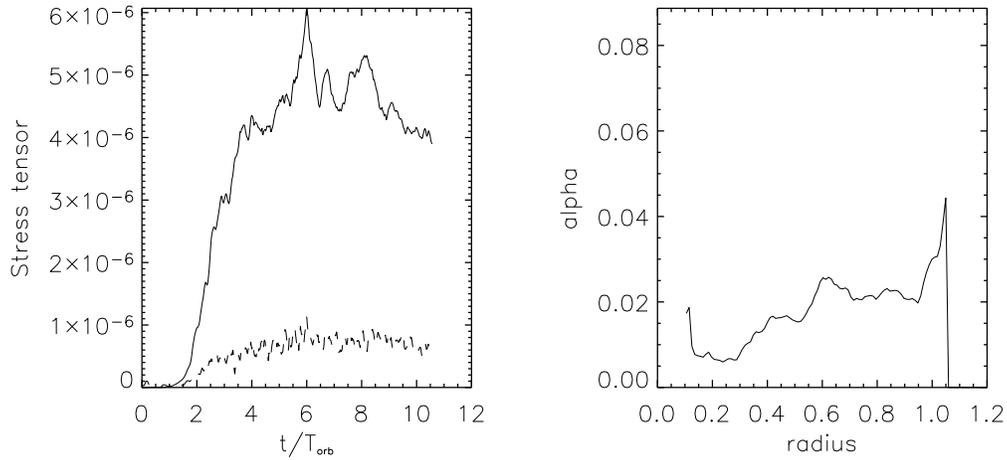}
\caption{{\em Left panel:} Time evolution of the volume averaged
Maxwell ({\it solid line}) and Reynolds ({\it dashed line}) stress
tensors for model T1. The Maxwell stress increases during the linear
growth of the MRI (first $4$ orbits).  It then saturates when the
instability breaks down into turbulence and stays roughly constant.
At all time, the Reynolds stress is much smaller than its magnetic
counterpart. {\em Right panel:} $\alpha$ vs. $r$ at the end of the
simulation, i.e. at $t \simeq 10$. The typical value of $\alpha$ is a
few times $10^{-2}$.}
\label{3d_evol_mhd_axi}
\end{center}
\end{figure*}

Figure~\ref{3d_evol_mhd_axi} shows the time evolution of the
volume-averaged Maxwell and Reynolds stress tensors and the
corresponding
$\alpha$ parameter (see eq.~[\ref{eq:alpha}]).  The Maxwell stress
increases during the 
linear phase of the instability.  It then saturates after
4~orbits, when the MRI breaks down into turbulence.  The presence of
turbulence is seen in figure~\ref{density_mhd_axi}, which shows the
density perturbation in the equatorial plane at $t=6.4$.  Turbulent
fluctuations are present over the full extent of the disk.  It is
clear from figure~\ref{3d_evol_mhd_axi} that the Reynolds stress is
significantly smaller than the Maxwell stress over the course of the
simulation.  This is in agreement with previous non self--gravitating
global simulations of the MRI
\citep{hawley00,hawley01,steinacker&pap02}.  The right panel of
figure~\ref{3d_evol_mhd_axi} shows the radial profile of $\alpha$ at
the end of the simulation, i.e. at $t \simeq 10$.  The typical value
of $\alpha$ is a few times $10^{-2}$, similar to what was found in
previous simulations starting with a toroidal field with a net flux
\citep{steinacker&pap02}.

\begin{figure}
\begin{center}
\epsscale{.45}
\vspace{6cm}
\caption{Density perturbation in the equatorial plane for run T1 at
$t=6.4$.  Turbulent fluctuations are present over the whole extent
of the disk.}
\label{density_mhd_axi}
\end{center}
\end{figure}

The Maxwell stress stays roughly constant during our simulation.
This indicates that the resolution $(128,32,128)$ is large enough for the
turbulence to be sustained over the duration of the run.  We therefore
adopt it in the following runs (which of necessity are limited in time
by the fact that mass is accreted onto the central mass).

\subsubsection{Initial poloidal field: Model P1}
\label{sec:P1}

To investigate the sensitivity of our results to the initial field
geometry, we run the same calculation as in model T1 but with an
initial poloidal magnetic field.  We calculate the field from the
(toroidal) component of the vector potential in the initial disk
model:

\begin{equation}
A_{\phi} \propto \rho \cos \left( 8 \pi
\frac{r-R_{in}}{R_{out}-R_{in}} \right) \, .
\label{Aphi}
\end{equation}

\noindent
This corresponds to 4~magnetic loops confined
inside the disk.  The first 2D simulations of a disk permeated by a
weak vertical field \citep{balbus&hawley91b} showed the development and
growth of ``channel'' solutions.  In 3D, these solutions still exist but
they quickly break down into turbulence, as predicted by the
analysis of \citet{goodman&xu94}.  Turbulence is more rapidly
established when the field varies on a fairly small scale, which 
motivates the above choice of $A_\phi$.

The radial and vertical components of the magnetic field are computed
from $A_{\phi}$ and normalized such as to obtain the desired initial
value of $\langle \beta \rangle$.  Since the linear growth of the
vertical field is much more rapid than that of the toroidal field (see
below), we chose a much larger initial value of
$\langle \beta \rangle=300$. 

\begin{figure}
\begin{center}
\epsscale{0.55}
\plotone{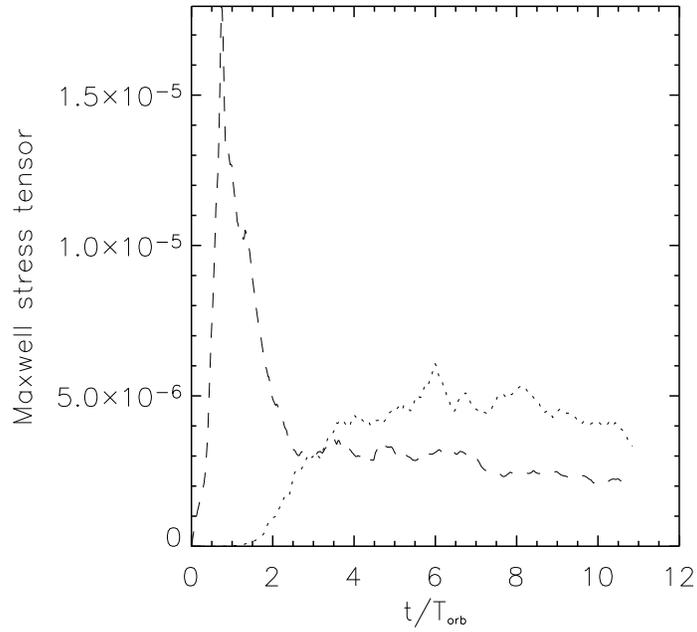}
\caption{Time history of the volume averaged Maxwell stress tensor for
runs P1 ({\it dashed line}) and T1 ({\it dotted line}).  The linear
instability is more vigorous when a vertical field is present, but the
level of turbulence is similar in both cases. }
\label{maxwell poloidal axi}
\end{center}
\end{figure}

The properties of the turbulence are similar to those found when the
initial field is toroidal.  Figure~\ref{maxwell poloidal axi} shows
the time evolution of the Maxwell stress for both models T1 and P1.
As expected, the linear instability is much more vigorous when a
vertical field is present, because of the growth of the channel
solutions.  However, in both cases the stress saturates at a similar
value and the level of turbulence is comparable.  The evolution of the
Maxwell stress in P1 is somewhat similar to what was obtained in paper
I.  The important difference is that in 3D, the stress saturates when
turbulence is established and does not decay with time, as it does in
2D.

\subsection{Full MHD simulations}

In this section we report the results of full 3D simulations including
the development of both self-gravitational and magnetic instabilities.
The resolution is $(N_r,N_{\phi},N_z)=(128,64,128)$ and the azimuthal
domain extends from 0 to $\pi$.

\subsubsection{Initial Toroidal Field: Models T2, T2$^*$ 
and T3}

In this sequence of models, we observe the simultaneous appearance of both
MHD turbulence and the $m=2$ spiral arm familiar from the hydrodynamical
calculation.  To better understand how angular momentum is transported
in the disk, we compare the time evolution of the different stresses
with those obtained in the models described in the previous section.

\begin{figure}
\begin{center}
\epsscale{.6}
\plotone{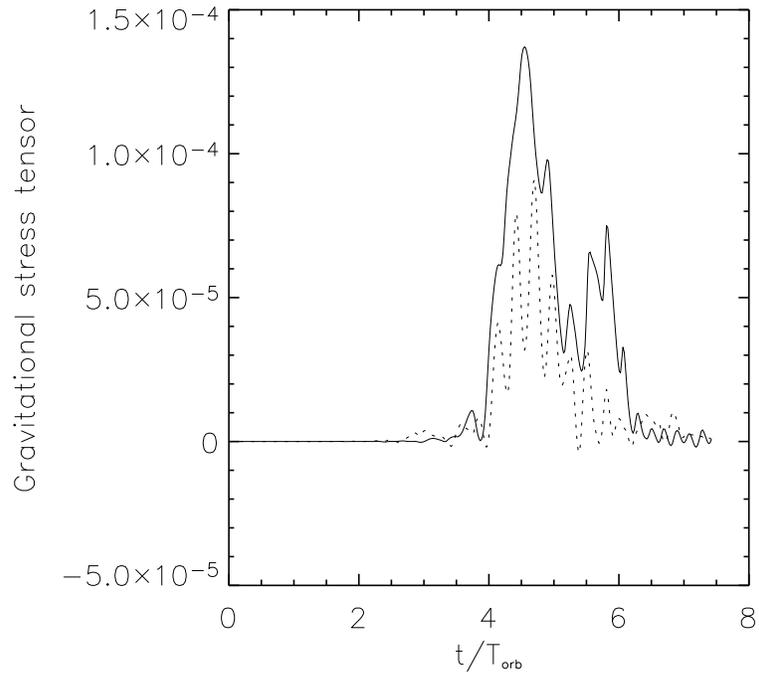}
\caption{Time evolution of the volume averaged gravitational stress
tensor $\langle T_{r\phi}^{grav} \rangle$ for models HD ({\it solid
line}) and T2 ({\it dotted line}).  The level of transport by
gravitational instabilities is significantly reduced when MHD
turbulence is present.  In model T2, the gravitational stress also
varies periodically.}
\label{grav toroidal}
\end{center}
\end{figure}

\begin{figure*}
\begin{center}
\epsscale{0.8}
\plottwo{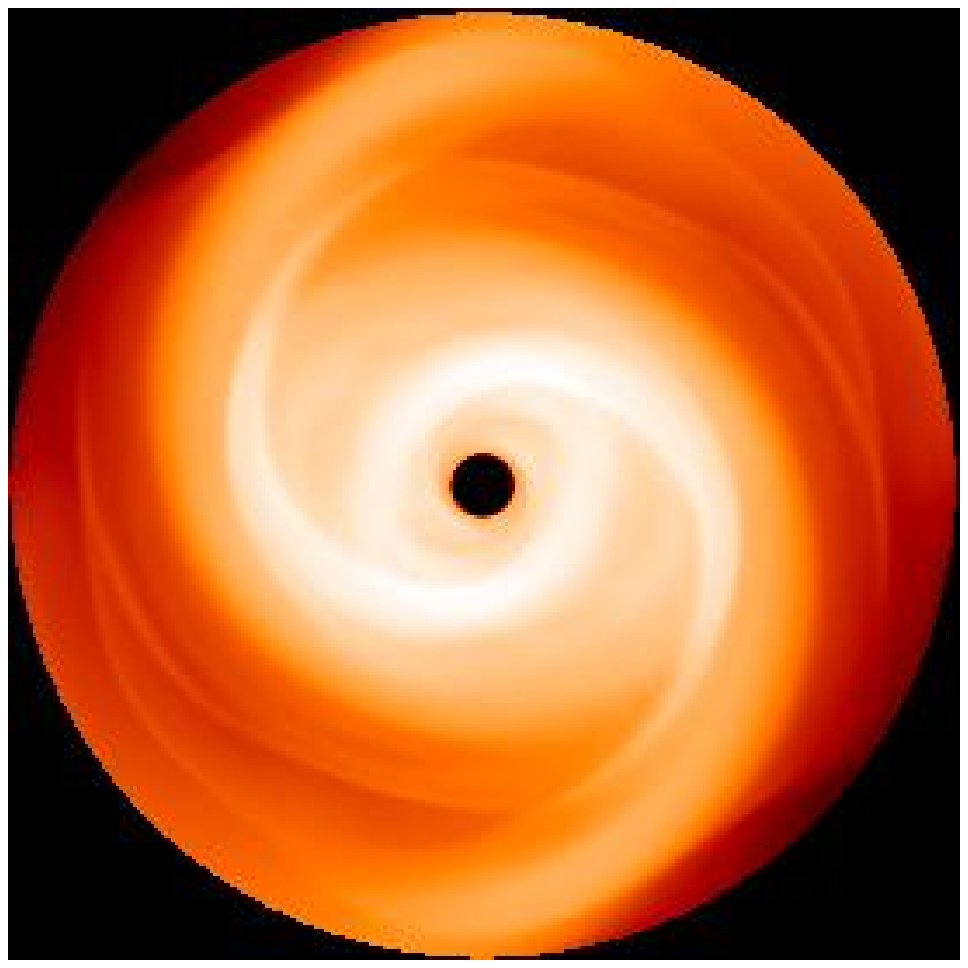}{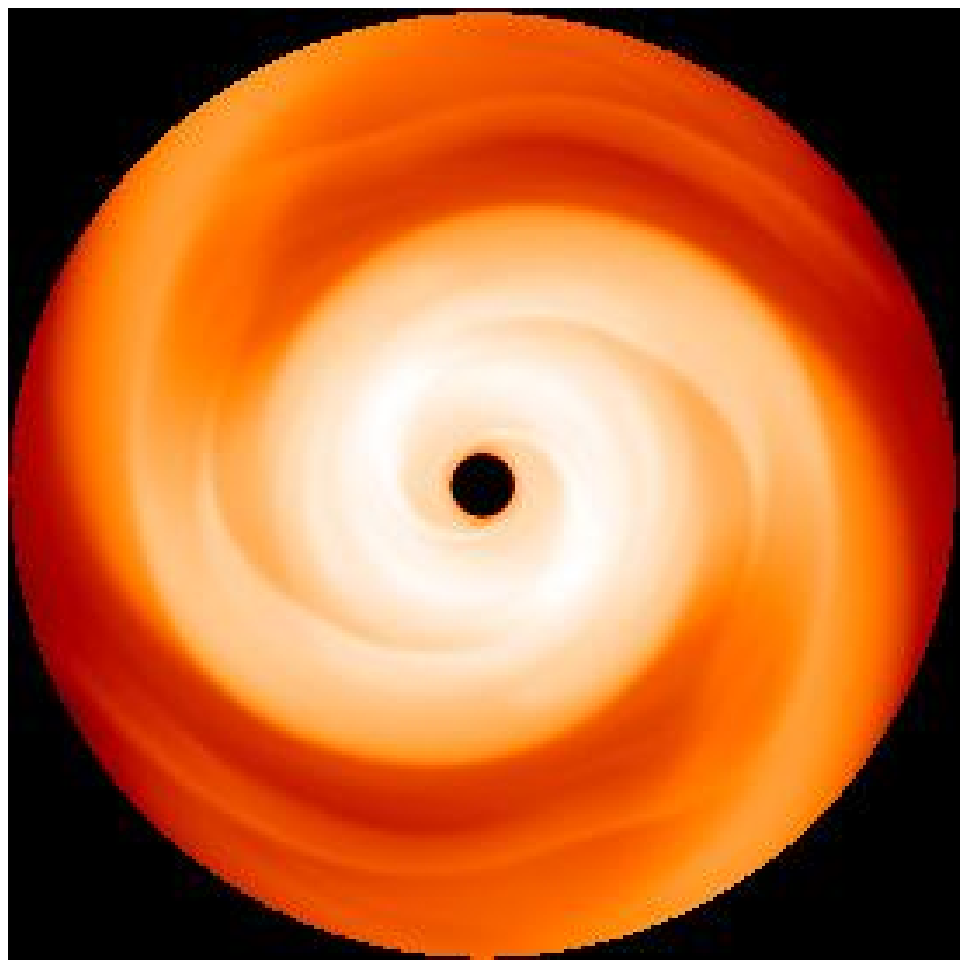}
\caption{Logarithm of the density in the equatorial plane for run T2
at $t=4.95$ ({\it left panel}) and $t=5.09$ ({\it right panel}).
These snapshots correspond to a maximum and a minimum of the
gravitational stress, respectively.  The spiral arms are sharp and
clear at $t=4.95$, whereas they appear blurred at $t=5.09$.}
\label{snapshot toroidal}
\end{center}
\end{figure*}

Figure~\ref{grav toroidal} shows the gravitational stress $\langle
T_{r\phi}^{grav} \rangle$ as a function of time for both models T2 and
HD.  Somewhat surprisingly, the presence of
both gravitational and MHD instabilities
leads to an average $\langle T_{r\phi}^{grav} \rangle$ reduced by a
factor $ \sim 2$ compared with the values obtained without a magnetic
field.  The magnetic torques do not lead to more vigorous
gravitational instability.  
One possible explanation may be that turbulent motions tend to
broaden the spiral arms by adding an extra fluctuating component to
the thermal pressure, but there is more going on
just this.  Figure~\ref{grav toroidal} also shows
that the gravitational stress varies nearly periodically with time and can
reach very small values.  This behavior is in fact associated with the
near disappearance of the spiral arms, as can be seen in
figure~\ref{snapshot toroidal}.  These snapshots correspond to a
maximum and a minimum of the gravitational stress, respectively.  The
spiral arms are sharp at $t=4.95$, whereas they lack definition at
$t=5.09$.  The arms form, disperse, and reform.
This periodic variation is also seen in figure~\ref{mdot
toroidal}, which shows the mass accretion rate onto the central mass
as a function of time.  As expected, the accretion rate has a periodic
component with the same frequency as that found in the gravitational
stress.  The period in both cases is $\sim 0.28$. 

\begin{figure}
\begin{center}
\epsscale{0.45}
\plotone{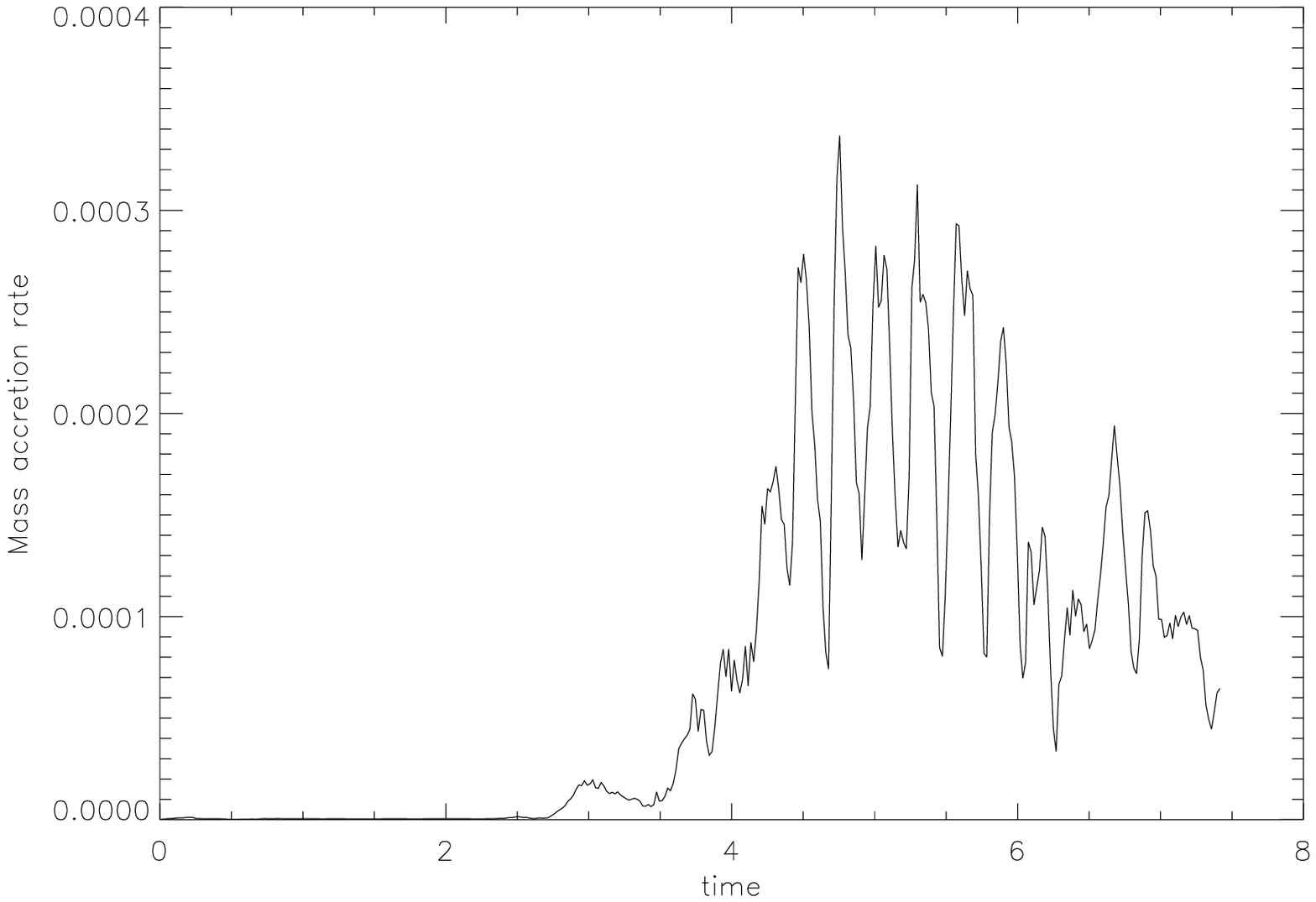}
\caption{Time evolution of the mass accretion rate (mass per unit of
time) onto the central mass for run T2.  As expected, the accretion
rate oscillates with the same period as the gravitational stress.}
\label{mdot toroidal}
\end{center}
\end{figure}

\begin{figure}
\begin{center}
\epsscale{0.7}
\plotone{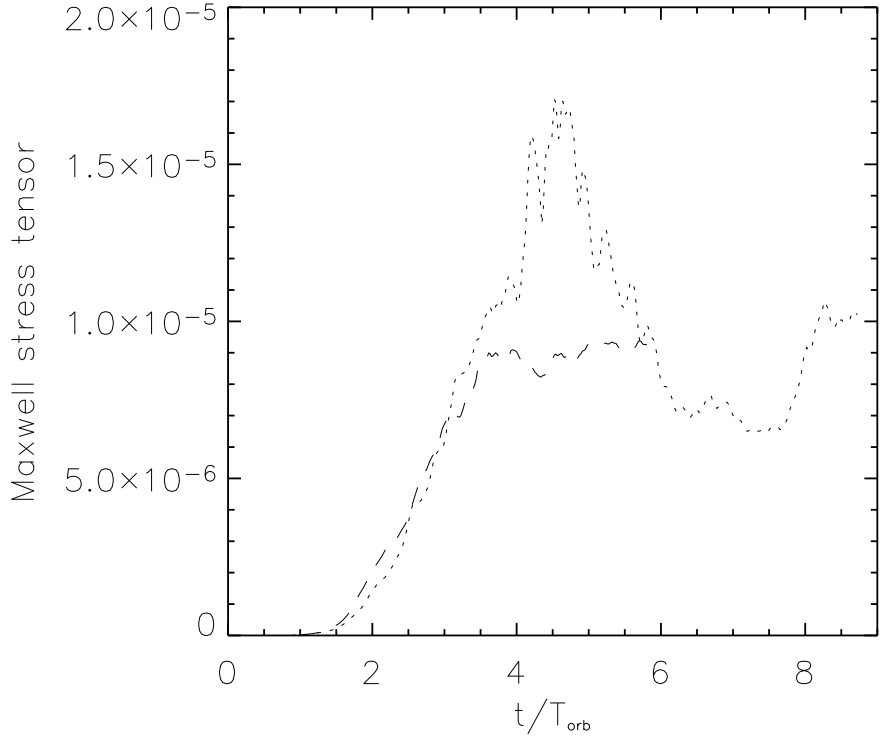}
\caption{Time evolution of the volume averaged Maxwell stress tensor
for runs T2$^*$ ({\it dashed line}), before gravitational transport is 
turned on, and T2 ({\it dotted line}).  The
Maxwell stress is larger when the disk is gravitationally unstable.
This appears to be due to a compression of the magnetic field lines
along the spiral arms.  When the disk in model T2 becomes
gravitationally stable (after $t \simeq 7$), the Maxwell stress
decreases down to the same value as in model T2$^*$.}
\label{maxwell toroidal}
\end{center}
\end{figure}

In figure~\ref{maxwell toroidal}, we compare the evolution of the
Maxwell stress in models T2 and T2$^*$ (for which only the first $6$
orbits, 
during which $m_{max}=0$, are plotted).  In contrast
to the gravitational stress, the Maxwell stress
is significantly larger when the disk is
gravitationally unstable.  This appears to be due to the systematic
compression of the magnetic field lines along the spiral arms, as
opposed, say, to an increase of the level of turbulent fluctuations.  When
the gravitational instability disappears after about $7$ orbits in
model T2, for example, the Maxwell stress decreases to the same value
as in model~T2$^*$.

As in the hydro model~HD, the Toomre $Q$ parameter rises
throughout the body of the disk over the course of the simulation 
(as mass is transported toward the inner region), until the
gravitational instability ceases. But even by $t \simeq 8$, when there is
no longer any gravitational transport, $Q$ is still larger in model~HD 
than in model~T2. This is because the gravitational instability is 
stronger in the hydrodynamical case, and the disk is depleted more rapidly.

To check the sensitivity of these results to our choice of initial
conditions, we conducted the following experiment.  In model T2$^*$, the
input parameters are the same as in model T2, but the non--axisymmetric
part of $\Phi_s$ is included only after 6~orbits, i.e. only after MHD
turbulence has been firmly established.  Figure~\ref{maxwell gravitational
toroidal} shows the time evolution of the volume averaged Maxwell and
gravitational stress tensors for run T2$^*$.  Until $t=7$, the development
of MHD turbulence is the same as in model T1.  However, in the time
interval $t=7$--8, i.e. after gravitational instabilities have developed,
$\langle T_{r\phi}^{Max} \rangle$ decreases to reach about one third of
its value at $t=7$.  The reason of this decline is not completely clear.
One possibility may be that the compression of the (randomized) magnetic
field in the spiral arms leads to more efficient reconnection of the field
lines.  Another possibility is that gravitational stresses feed off the
density fluctuations generated by the MRI, thereby indirectly coupling
the magnetic and gravitational energies.  In any case, this behavior
stands in contrast with was observed in run~T2, where gravitational
instabilities developed while the magnetic field was still ordered.
In model T2$^*$, the gravitational stress tensor is roughly a factor of
$2$ smaller than in model T2, but shows the same periodic variations.

\begin{figure}
\begin{center}
\epsscale{0.7}
\plotone{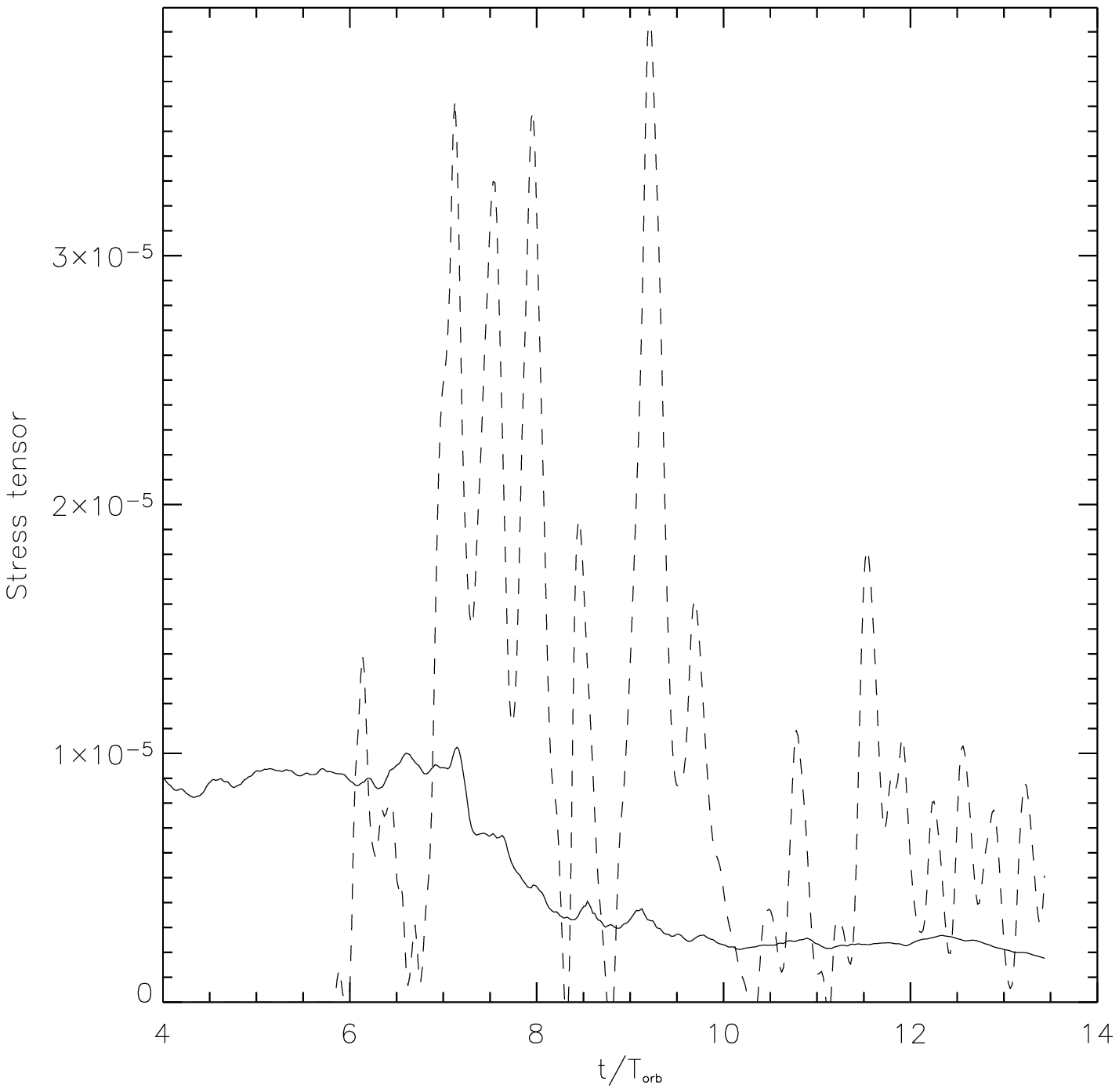}
\caption{Time evolution of the volume averaged Maxwell ({\em solid
line}) and gravitational ({\em dashed line}) stress tensors for run
T2$^*$.  In this run, the non--axisymmetric part of $\Phi_s$ is
included only after 6~orbits, i.e. only after MHD turbulence has been
firmly established.  The development of gravitational instabilities
coincides with a significant decrease of the Maxwell stress tensor.  This
may be due to reconnection of the (randomly oriented) field lines in
the spiral arms.  The gravitational stress tensor has the same
amplitude and periodic variations as in run T2.}
\label{maxwell gravitational toroidal}
\end{center}
\end{figure}

Our next comparison run, model T3, differs from model T2 only in the
number $m_{max}$ of fourier coefficients in the expansion of $\Phi_s$.
(T3 has $m_{max}=16$, T2 has $m_{max}= 8$.) Once again, very similar
results emerge, and the choice of $m_{max}$ does not appear to be
critical (cf.\ \S~\ref{modal analysis}).

To summarize: the evolution of a purely toroidal field in a
gravitationally unstable disk leads to a reduction and strong periodicity
in the gravitational stress (compared with a purely hydrodynamical model).
Compared with gravitationally stable models, the Maxwell stress is larger
or smaller depending repsectively on whether gravitational instabilities
develop at the same time as MHD turbulence (magnetic field alignment in
spiral arms) or after turbulence is established (reduction in magnetic
stress as gravitational stress develops).  The behavior of an initial
poloidal field is considered next.

\subsubsection{Initial Poloidal Field: Model P2}

Do our toroidal field findings extend to poloidal field behavior?  To
answer this question, we begin with an initial poloidal field,
calculated as in section~\ref{sec:P1} above.  Again, we start with
$\langle \beta \rangle = 300$.  Except for the initial field geometry,
model P2 is the same as model T2.

\begin{figure}
\begin{center}
\epsscale{0.5}
\plotone{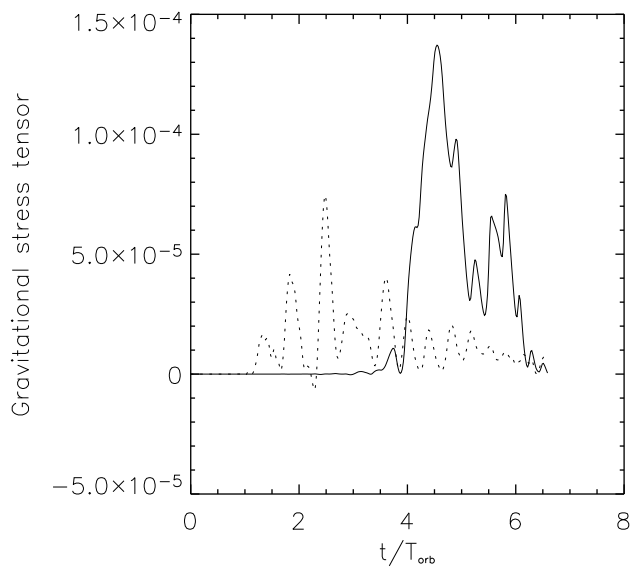}
\caption{Time evolution of the volume averaged gravitational stress
tensor for models HD ({\it solid line}) and P2 ({\it dotted line}).
The gravitational stress in model P2 is smaller than in model HD and
varies periodically with time.  }
\label{grav poloidal}
\end{center}
\end{figure}

Figure~\ref{grav poloidal} shows the evolution of the gravitational
stress tensor for both models P2 and HD (this is the equivalent of
figure~\ref{grav toroidal}).  As in the case of a toroidal field, a
non--axisymmetric $m=2$ spiral grows and becomes nonlinear in model
P2.  The fact that gravitational instabilities develop earlier in
model P2 than in models T2 and HD (see figures~\ref{grav toroidal} and
\ref{grav poloidal}) appears to be due to the fact that the strong linear
magnetic instability associated with the poloidal field produces large
perturbations of the density.  Once again, we find that the
gravitational stress is smaller than in model HD, and varies
periodically with time with a period $\sim 0.38$ somewhat larger than
for model P2.

\begin{figure}
\begin{center}
\epsscale{0.5}
\plotone{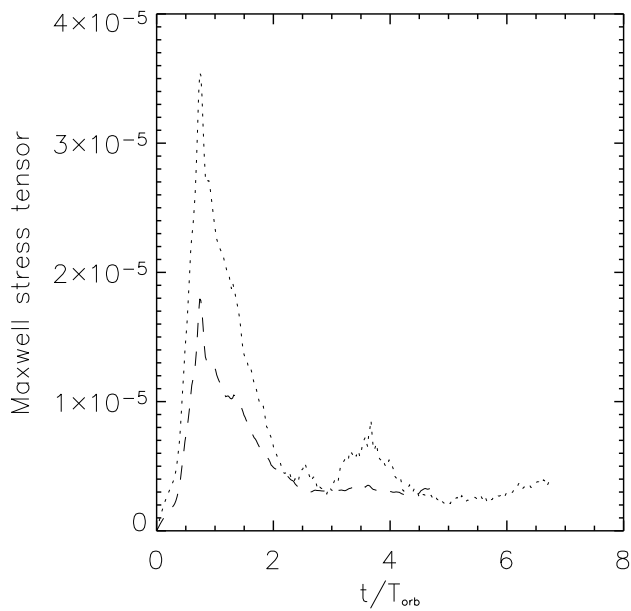}
\caption{Time evolution of the volume averaged Maxwell stress tensor
for models P1 ({\it dashed line}) and P2 ({\it dotted line}). The
Maxwell stress is larger when the disk is gravitationally unstable and
gravitational instabilities develop at the beginning of the run.}
\label{maxwell poloidal}
\end{center}
\end{figure}

Figure~\ref{maxwell poloidal} shows the evolution of the Maxwell
stress for models P1 and P2 (this is the equivalent of
figure~\ref{maxwell toroidal}).  Once again, the Maxwell stress is
larger when the disk is gravitationally unstable and gravitational
instabilities develop at the beginning of the run (model P2).

Since the state of MHD turbulence in a saturated disk is the same whether
an initial poloidal or toroidal field is used, we have not run a case
``P2$^*$'' with the non--axisymmetric part of $\Phi_s$ added later.
Such a run is expected to be very similar to T2$^*$, since the
initial turbulent states are similar.

The history of run P2 and the above argument together suggest 
that toroidal and poloidal initial fields behave very similarly.

\subsubsection{Modal analysis}
\label{modal analysis}

The full MHD simulations described above suggest that a general
feature of the evolution of gravitationally unstable turbulent disks
is a periodic modulation of the gravitational stress.  To understand
the reason for this modulation, we now analyse in more detail the
unstable modes that appear in models HD, T2 and P2.  We are in
particular interested in determining the power spectrum as a function
of mode frequency $\sigma$.  Following \citet{pap&savonije91}, we
Fourier
transform in time at each radial zone $r$ and at a fixed azimuth
$\phi_0$ the function $\rho(r,\phi_0,t)$.  To get the spectral time
evolution, we carry out each Fourier transform over a series of 4
distinct time intervals.  The number of time-steps used in each time
interval gives a finite frequency resolution $d \sigma / 2 \pi=0.3$.
The contours of constant power are then plotted as a function of
frequency and radius for the various time intervals.

\begin{figure}
\begin{center}
\epsscale{1.}
\plottwo{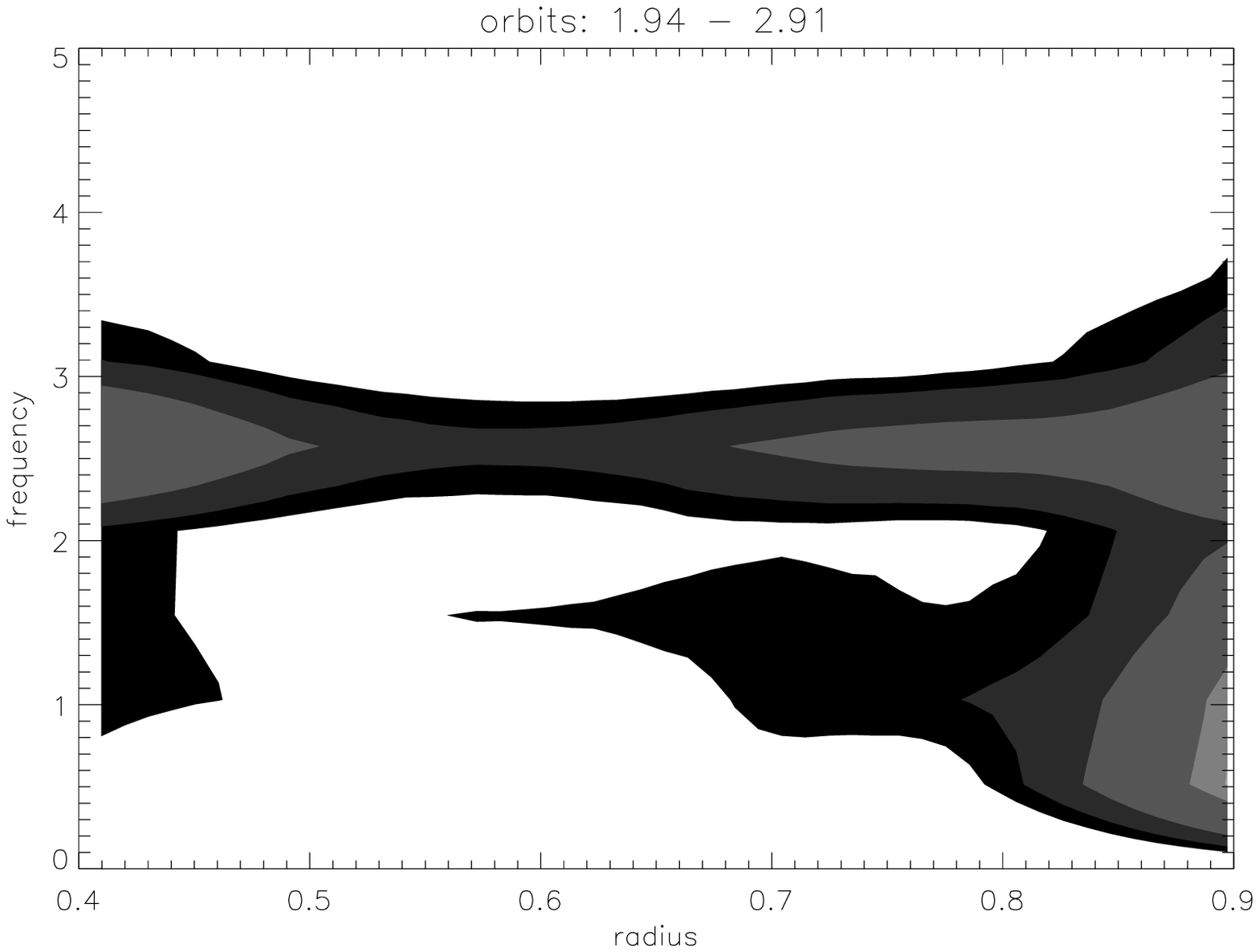}{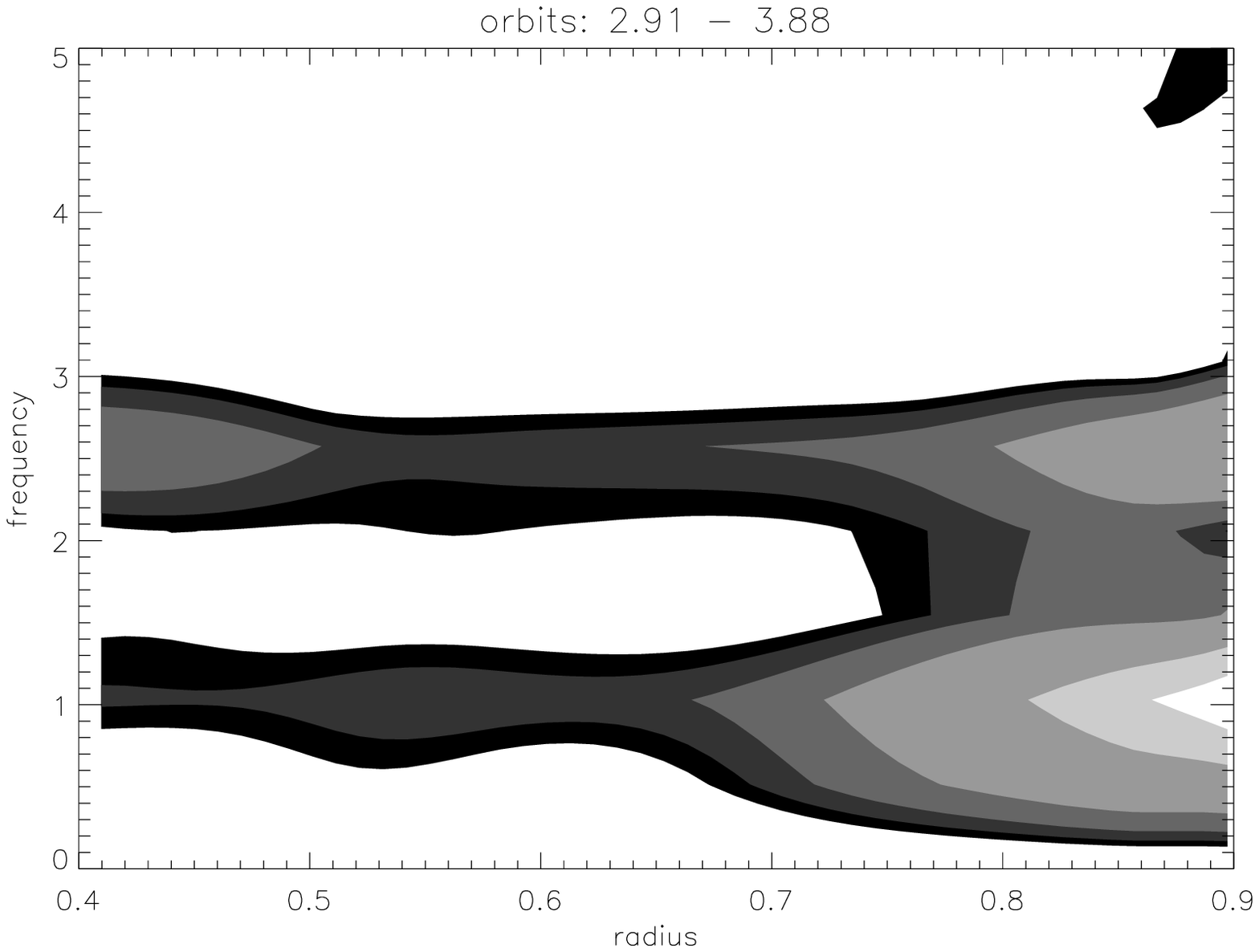} \\
\plottwo{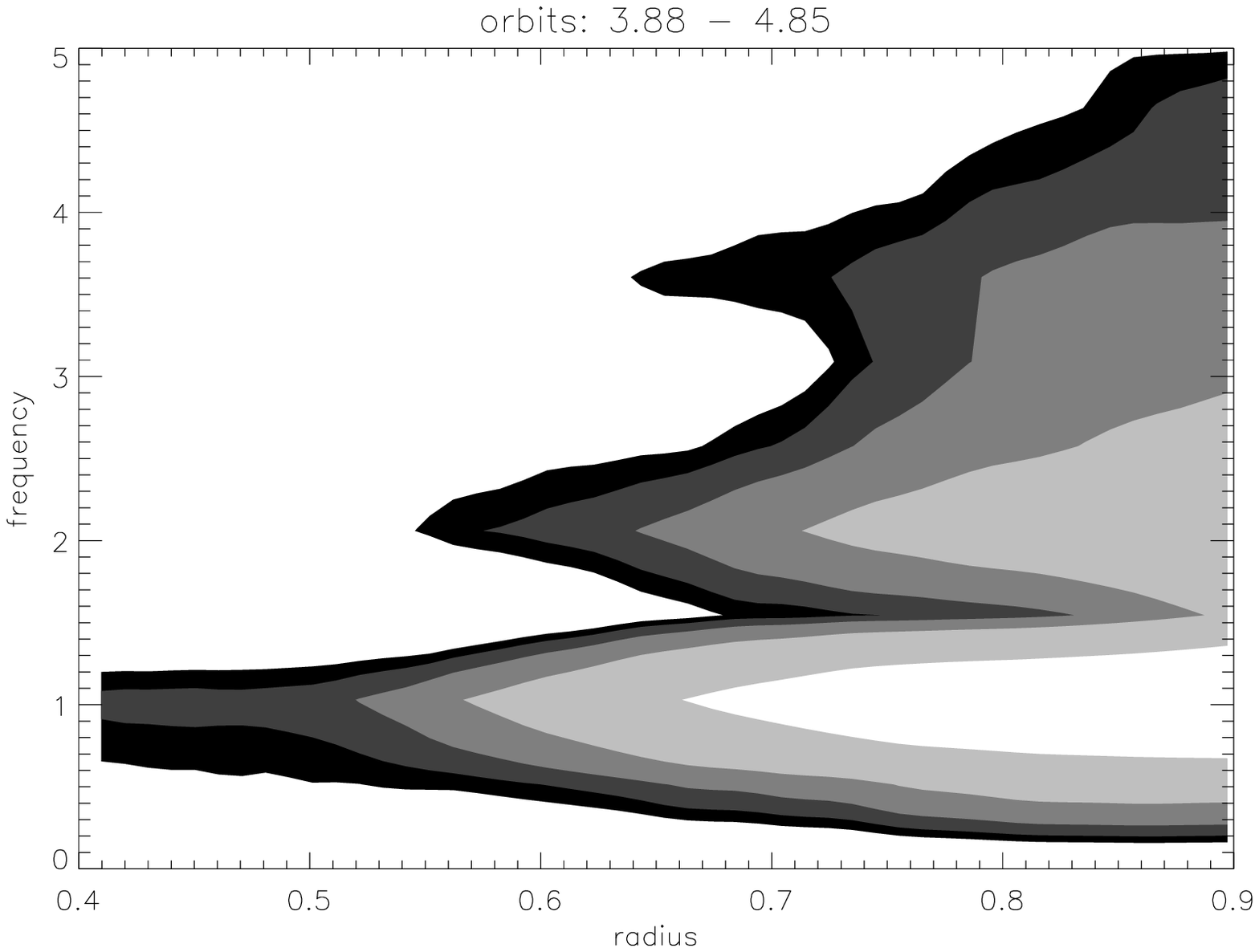}{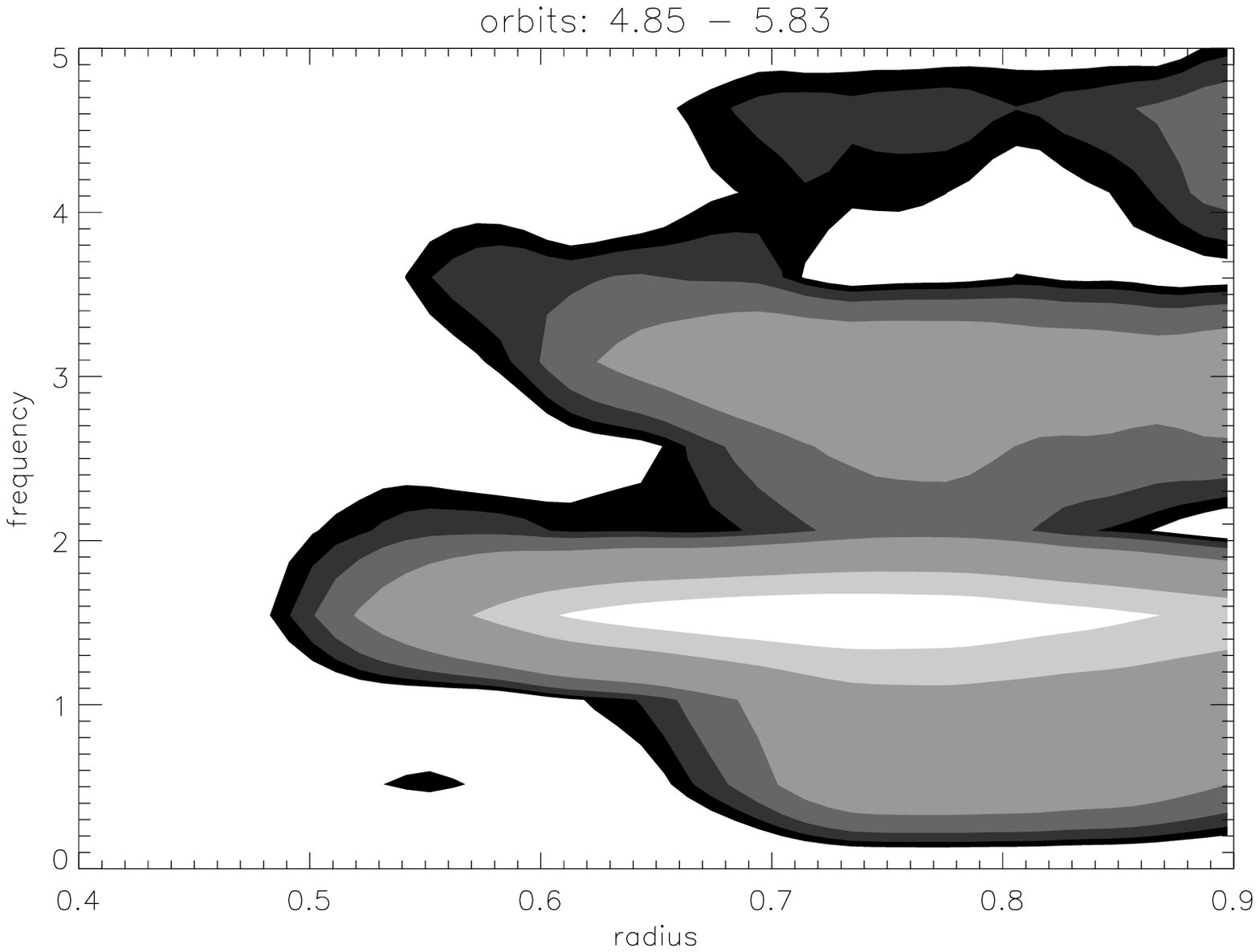}
\caption{Contours of constant Fourier power as a function of
dimensionless mode frequency $\sigma/ 2\pi$ ({\em vertical axis}) and
radius $r$ ({\em horizontal axis}) for model HD.  The different panels
correspond to the time intervals 1.94--2.91 ({\em upper left}),
2.91--3.88 ({\em upper right}), 3.88--4.85 ({\em lower left}) and
4.85--5.83 ({\em lower right}). The contour levels in each of these
panels are, in unit $10^{-2}$: (0.22; 0.29; 0.44; 0.87; 1.1; 1.6;
3.2), (2.1; 2.6; 3.4; 5.2; 10; 15), (8.4; 11; 14; 21; 42) and (9.4;
12; 16; 23; 47; 62), respectively.  One mode (a two--arm spiral)
dominates the spectrum in the non--linear stage (third panel), with a
frequency $\sigma / 2 \pi=1$.}
\label{mode hydro}
\end{center}
\end{figure}

Figure~\ref{mode hydro} shows the contours for model~HD.  The
different panels correspond to different time intervals.  In the first
panel, i.e. at time $t \simeq 2.4$, we see the presence of a mode with
frequency $\sigma/ 2\pi=2.5$ which extends over the whole disk.  Its
amplitude is small, peaking at about $3 \times 10^{-2}$.  This mode is
still present in the second panel, at time $t \simeq 3.4$, with a
similar amplitude structure.  However, a second low frequency
mode with $\sigma/ 2\pi=1$ (i.e. with a corotation radius at the disk
initial outer edge) is now apparent.  Its amplitude is significant
only in the disk outer parts, where it peaks at 0.2.  This mode
subsequently grows and completely dominates the high frequency
($\sigma / 2\pi=2.5$) mode in the third panel, i.e. at $t \simeq 4.4$.
There its amplitude peaks at 0.7.  This mode can of course be
identified with the two--arm spiral seen in figure~\ref{image220
hydro}.  At $t=4.4$ the mode is non--linear.  Its amplitude does not
increase further, as can be seen in the fourth and last panel, i.e. at
$t \simeq 5.4$.  At this later time, only the frequency of the mode
has changed.  It is now around 1.5.  Both the finite frequency
resolution $d \sigma$ and the increase of the central mass due to
accretion may account for this shift in frequency.  A third mode with
a frequency twice that of the low frequency mode is seen in the last
two panels.  The relationship between the frequencies of these two
modes and the fact that their radial structure is very similar
suggests that they are harmonics of each other.

\begin{figure}
\begin{center}
\epsscale{1}
\plottwo{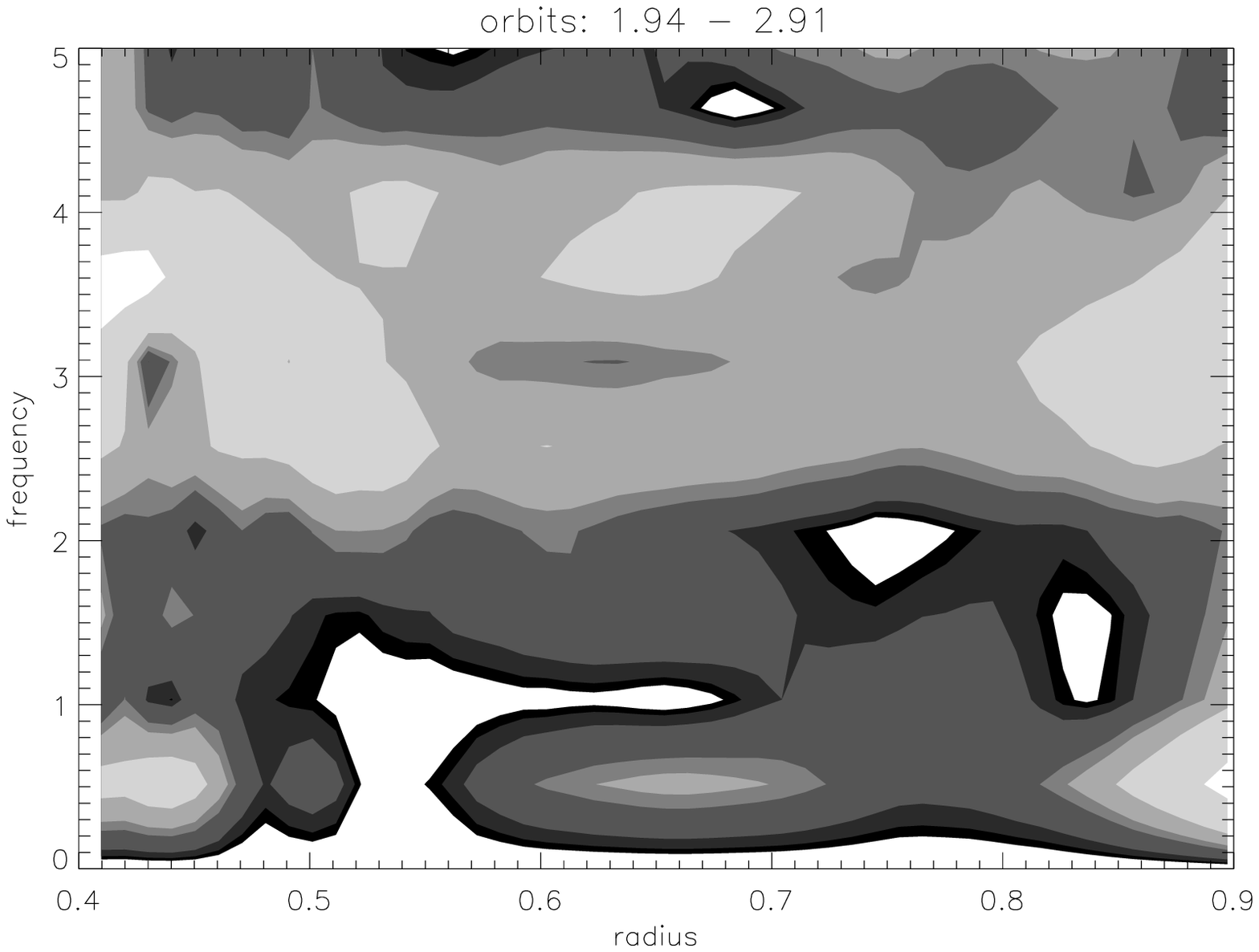}{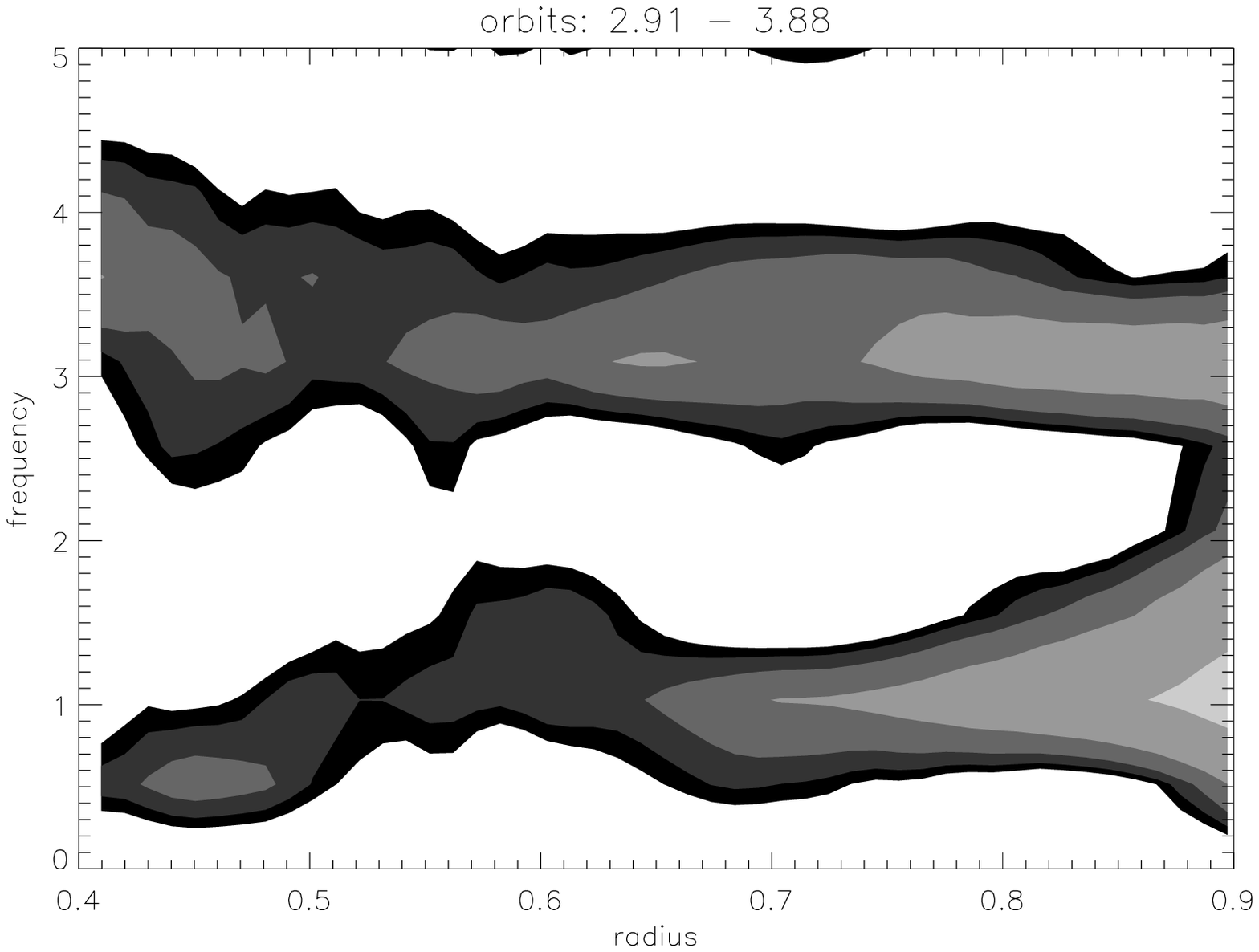}
\plottwo{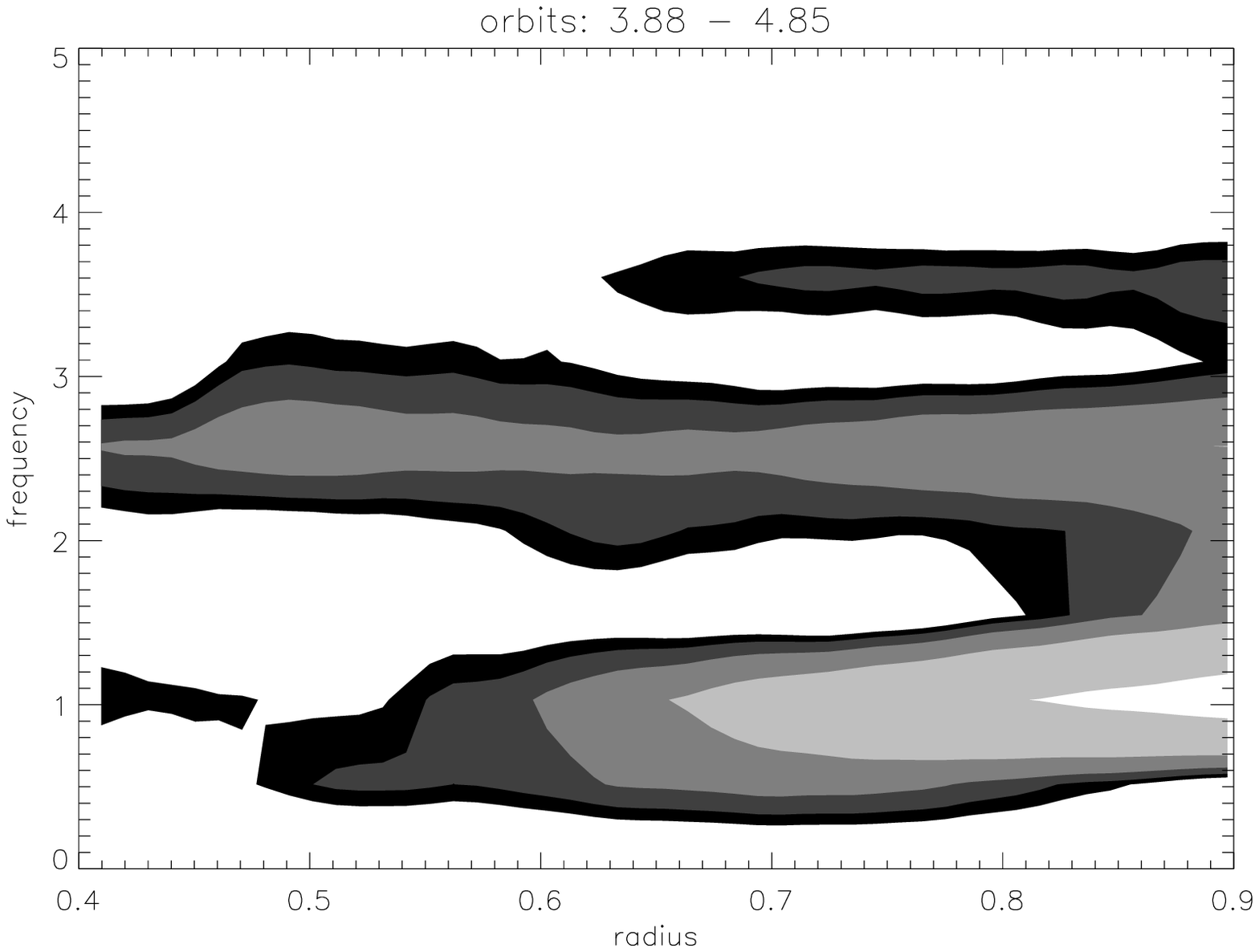}{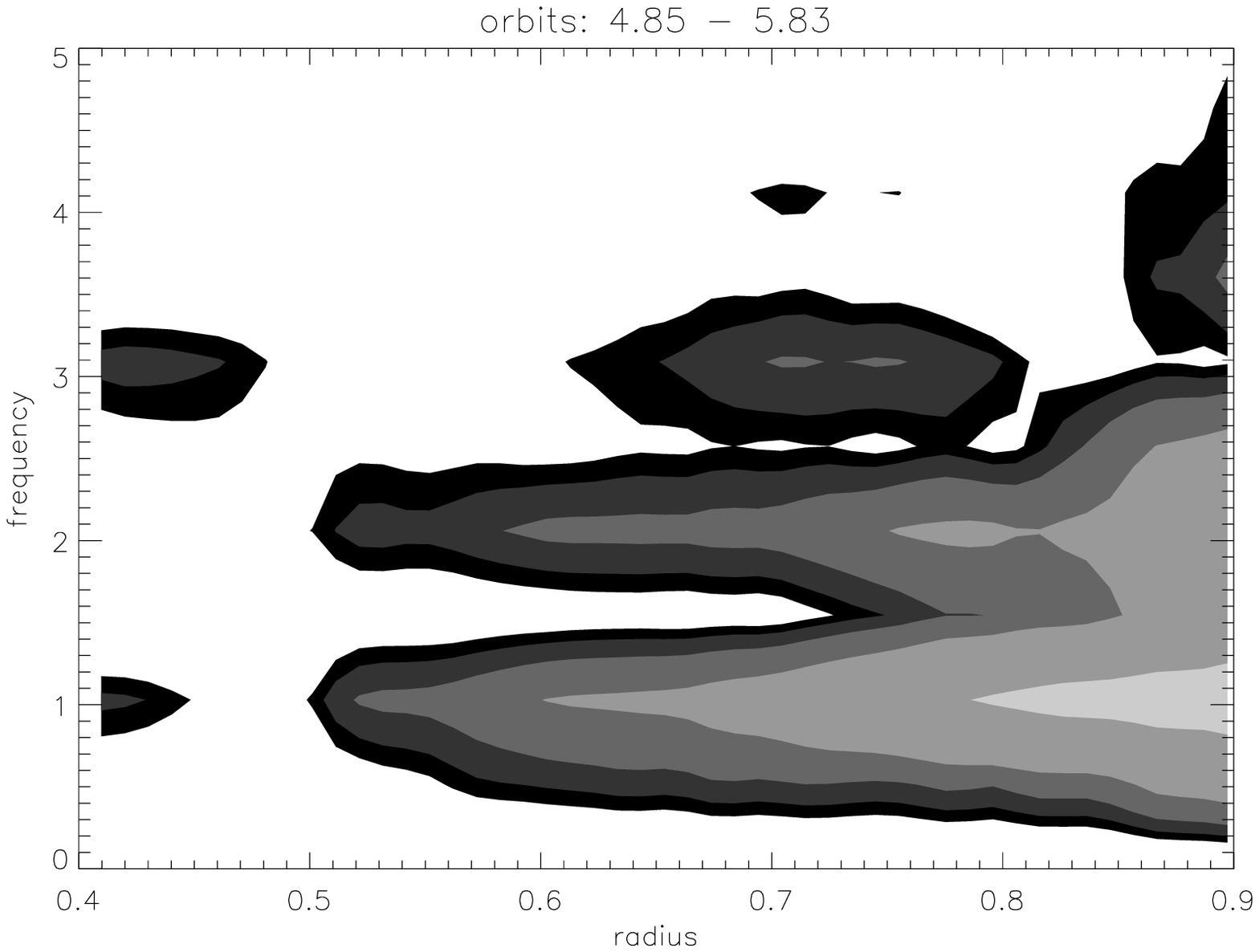}
\caption{Same as figure~\ref{mode hydro} but for model T2.  The
contour levels in each panel are the same as in figure~\ref{mode
hydro}.  Here two modes dominate the spectrum in the nonlinear phase
(third panel), with frequencies $\sigma/ 2 \pi=1$ and $\sigma / 2
\pi=2.5$,
respectively.}
\label{mode mhd}
\end{center}
\end{figure}

Figure~\ref{mode mhd} shows the contour plots for model~T2.  In the
first panel, at $t \simeq 2.4$, there is no dominant mode.  Instead,
there is a large number of high frequency (with mostly $\sigma/ 2
\pi=1.3$--3.8) perturbations.  The amplitude of these fluctuations is
a few times $10^{-2}$. They are associated with the growing MRI.  In
the second panel, at $t \simeq 3.4$, two modes emerge, but their
amplitude is still rather low.  However, these modes subsequently grow
and are clearly seen with a larger amplitude in the third panel, at $t
\simeq 4.4$.  One of these modes has a frequency $\sigma / 2 \pi=1$
and is the same as that seen in the hydrodynamical simulations.  Its
corotation radius is located at the disk initial outer edge.  Its
amplitude peaks at a value of about~0.5 in the outer parts of the
disk.  The other mode has a frequency $\sigma/ 2 \pi=2.5$ and an
amplitude $\sim 0.2$ constant over the whole disk.  In particular, its
amplitude in the disk inner parts is larger than that of the other
mode.  This mode is probably the same as that seen with a lower
amplitude at early times in the hydrodynamical simulations (panels~1
and~2 of figure~\ref{mode hydro}).  This suggests that this mode is a
disk eigenmode which is excited in model~T2 by the high frequency
motions associated with the turbulence.  We expect nonlinear coupling
between these two modes to give rise to beat oscillations,
i.e. oscillations with a frequency being a linear combination of the
frequencies of the two modes.  We have noted above that the
gravitational stress tensor in model~T2 oscillates with a period $\sim
0.28$.  This is consistent with the frequency of the oscillations
being $\sigma/2 \pi \simeq m( \sigma_{\rm HF} - \sigma_{\rm BF}) =3$,
where $\sigma_{\rm HF}$ and $\sigma_{\rm BF}$ are the frequencies of
the high and low--frequency modes, respectively.  This suggests that
the oscillations of the gravitational stress tensor result from a
nonlinear coupling between these two modes.

\begin{figure}
\begin{center}
\epsscale{.45}
\plotone{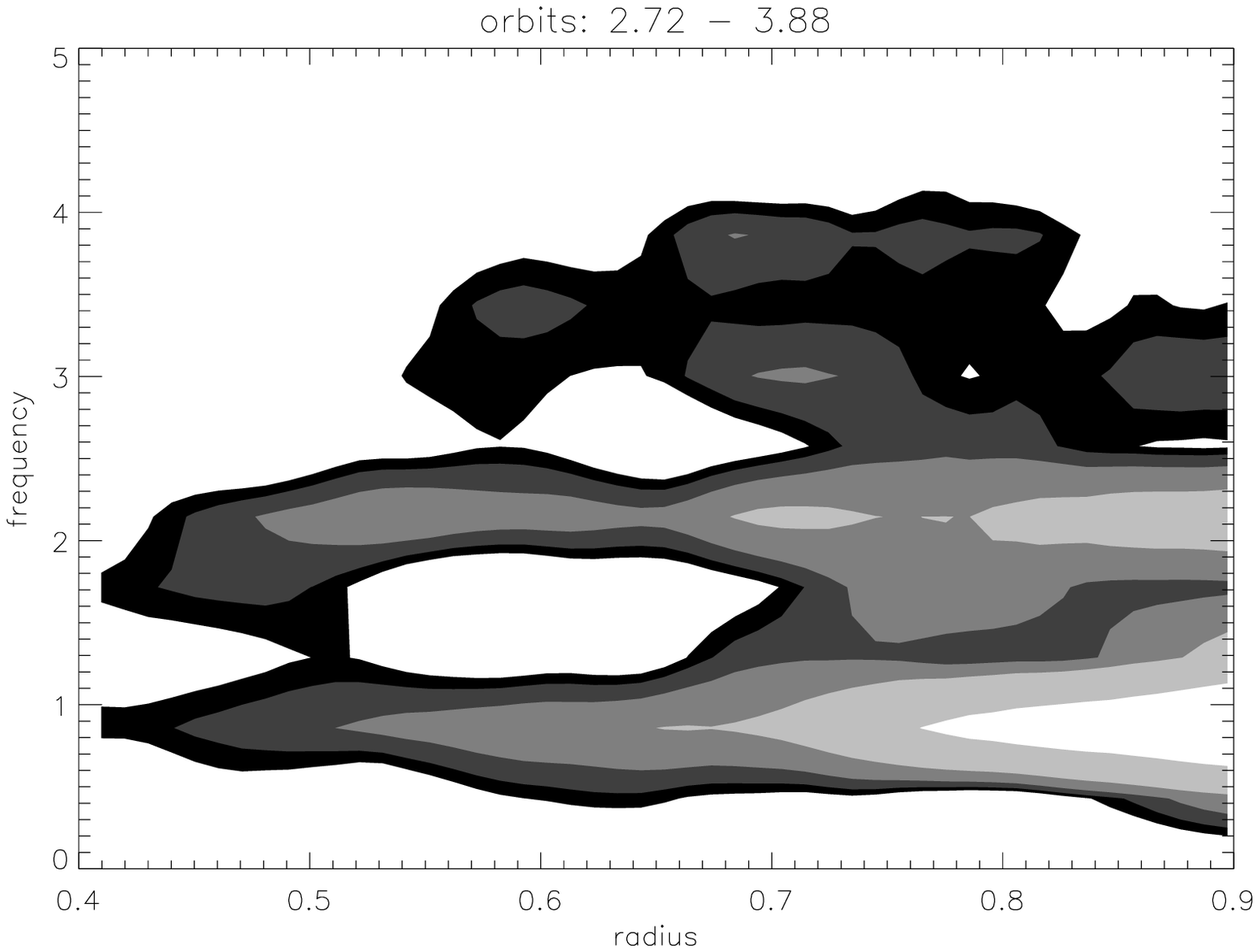}
\caption{Same as figure \ref{mode hydro} but for model P2 and only in
the time interval 2.72--3.88. The contour levels used here are, in
unit $10^{-2}$: (5.1; 6.4; 8.5; 13; 25).}
\label{mode poloidal}
\end{center}
\end{figure}

Figure~\ref{mode poloidal} shows the contour plots for model~P2 in the
time interval 2.72--3.88.  the situation is similar to model~T2.  Here
again two modes of comparable amplitude are present.  One of this mode
has a frequency $\sigma/2 \pi=1$ and can be identified with the mode
which emerges in the hydrodynamical simulations.  The other mode has a
frequency $\sigma/ 2 \pi=2.1$.  Given the finite frequency resolution $d
\sigma/ 2 \pi$, this second mode may be the same as that identified in
model~T2.  However, it may also be a different mode with a lower
frequency.  In any case, the situation is qualitatively the same here
as in model~T2.  Since the two modes have comparable amplitude, they
interact nonlinearly, which results in a periodic modulation of the
gravitational stress.

\begin{figure}
\begin{center}
\epsscale{.45}
\plotone{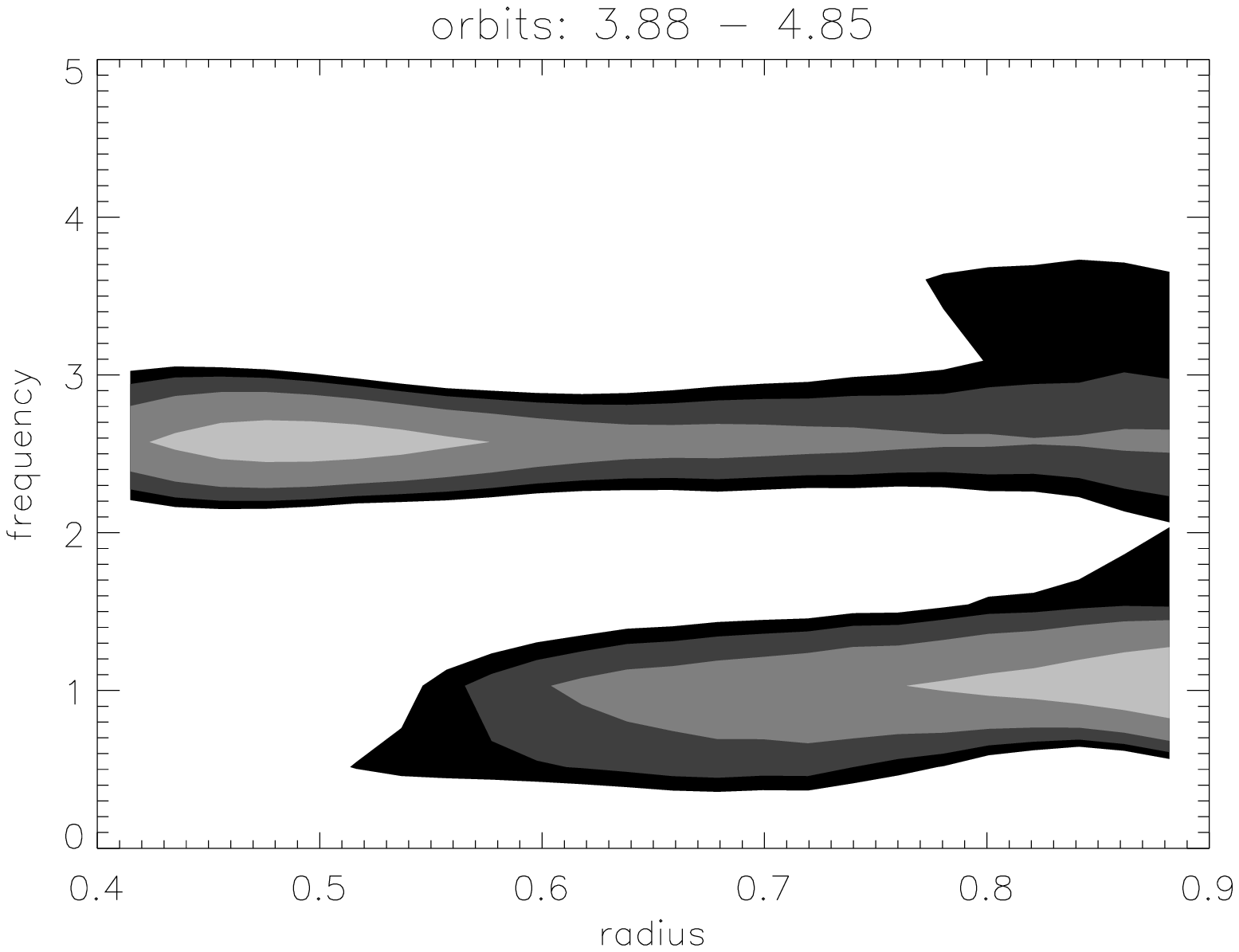}
\caption{Same as figure \ref{mode hydro} but for model T2$_{low}$ and only 
in the time interval 2.72--3.88. The contour levels used here are, in
unit $10^{-2}$: (5.1; 6.4; 8.5; 13; 25).}
\label{mode t2low}
\end{center}
\end{figure}

\begin{figure}
\begin{center}
\epsscale{.45}
\plotone{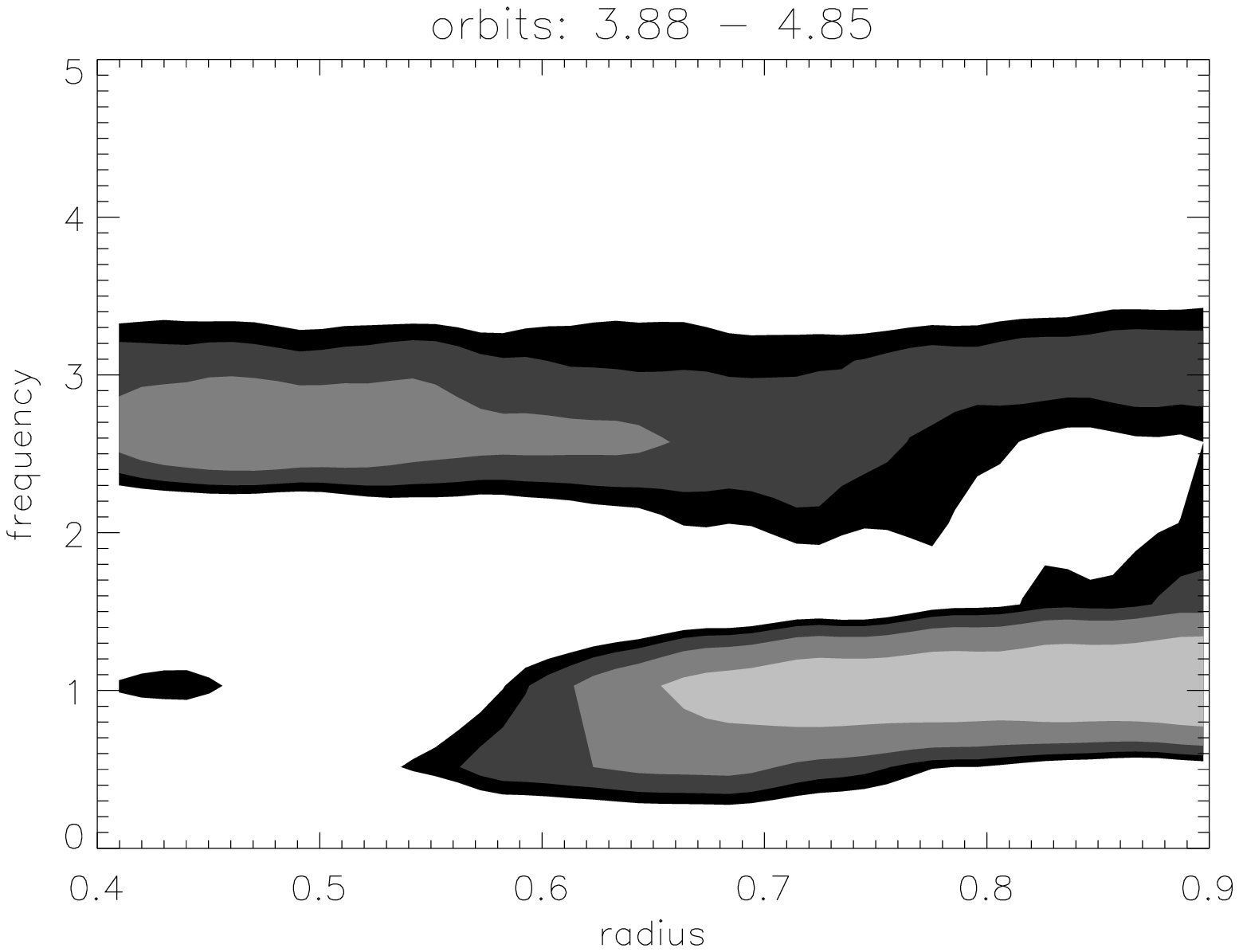}
\caption{Same as figure \ref{mode hydro} but for model T3 and only in
the time interval 2.72--3.88. The contour levels used here are, in
unit $10^{-2}$: (5.1; 6.4; 8.5; 13; 25).}
\label{mode t3}
\end{center}
\end{figure}

We now vary some of the parameters of model~T2 in order to examine the 
sensitivity of these physical results to the numerical input parameters.
Figure~\ref{mode t2low} shows the contour plots for model~T2$_{low}$ 
in the time interval 2.72--3.88. The resolution for this run in the radial 
and vertical directions is half that of model~T2. The 
similarity between this plot and the third panel of figure~\ref{mode mhd} 
demonstrates that our results do not depend strongly on the numerical 
resolution. Figure~\ref{mode t3} is the same as figure~\ref{mode t2low} but 
for model~T3. Again, it is very similar to the third panel of 
figure~\ref{mode mhd}. Since, in model~T3, only the parameter $m_{max}$ 
is different from model~T2, figure~\ref{mode t3} suggests that the limited 
number of Fourier components included in the calculation of the 
self--gravitating potential does not qualitatively affect the
main physical results 
presented in this paper. 

In conclusion, models P2, T2$_{low}$ and T3, taken together, suggest that the 
physical results described in this paper are insensitive to both
the numerical setup of the simulations and the initial magnetic 
field topology.

\section{Discussion}

In this paper, we have presented the first 3D numerical simulations of
the evolution of self--gravitating and magnetized disks.  We have
investigated disks in which only gravitational or magnetic
instabilities develop, and disks in which both types of instabilities
occur.

When no magnetic field is present, self-gravitating disks are unstable
when the Toomre $Q$ parameter is close to unity.  The spectrum of
unstable modes in that case is dominated by a large scale two--arm
spiral whose corotation radius is located near the disk outer edge.
The instability is due to the interaction between waves which
propagate near the disk outer boundary and inside the inner Lindblad
resonance (ILR), respectively \citep{pap&savonije91}.

When a magnetic field is present, two large scale modes grow in the
disk.  They both have $m=2$.  One of this mode is the same as that
seen in the hydrodynamical simulations.  The second mode has a higher
frequency.  The ILR of the low--frequency mode, which is at $r \simeq
0.6$, is very close to the corotation radius of the high--frequency
mode, which is at $r \simeq 0.5$.  Such a pair of modes was seen in
the hydrodynamical calculations of self--gravitating disks performed
by \citet{pap&savonije91}.  There, both modes were unstable
because of the particular vortensity profile.  In our simulations, in
the absence of magnetic field, only the low--frequency mode is
unstable.  The high--frequency mode seems to be part of the spectrum
of normal modes in that case, but it does not grow.  The presence of
MHD turbulence does not modify the spectrum of large scale (comparable
to the disk radius) modes, as it acts on scales limited by the disk
thickness.  However, the fact that the high--frequency mode is
unstable in the MHD simulations suggests that turbulence acts as a
source for high--frequency oscillations.

Nonlinear coupling between these two modes leads to an oscillation of
the gravitational stress tensor.  Note that such a coupling between
two modes with coinciding resonances has been suggested to explain
some of the features seen in numerical simulations of galactic disks
by \citet{taggeretal87}.  These authors argued that the proximity of
the resonances made the coupling very efficient.  The oscillation of
the gravitational stress tensor is accompagnied by the periodic
disappearance of the spiral arms in the disk.  Also, the peak value of
this stress is decreased by about half compared to the hydrodynamical
simulations.

The results reported here are robust and do not depend on the geometry
of the magnetic field.  They have important consequences for disks
around AGN and protoplanetary disks.  They first show that accretion
of a self--gravitating disk onto the central star is slowed down when a
magnetic field is present.  They also show that the accretion is
time--dependent, with a characteristic timescale for the variability
being on the order of a fraction of the dynamical timescale at the
outer edge of the region where the instabilities develop.

As mentioned in the introduction, protoplanetary disks are probably
self--gravitating in the early phases of their evolution.  For a disk of
about 100~AU, the work presented here suggests variability on a timescale
$\sim 10^3$~years.  The periodicity in the spatial distribution of knots
in jets emanating from such objects is in the range 10--$10^3$~years
\citep{reipurth00}, and is usually thought of as being produced by
a time--dependent accretion in the central parts of the disk. The
simulations presented in this paper suggest that periodic modulations
of the accretion rate might well be the result of the interplay between
gravitational instabilities and MHD turbulence, a far from obvious source.
Note that the first detection of near--IR variability in a sample of
Class~I protostars was performed recently by Park \& Kenyon (2002).
However, the poor time coverage of their data prevents a useful measure
of the variability timescales to be extracted.

Disks around AGN display time--dependent phenomena on a large range of
timescales \citep{ulrichetal97}.  The dynamical timescale for a disk
orbiting a $10^8$ solar masses black hole at $10^{-2}$ parsecs is
9.3~years, and variations are observed on timescales up to years.
This again is consistent with the processes described in this paper.

\section*{Acknowledgments}

It is a pleasure to thank John Hawley, John Papaloizou and Michel
Tagger for useful discussions.  SF thanks the Department of Astronomy
at UVa for hospitality during the course of this work.  SF is
supported by a scholarship from the French {\em Minist\`ere de
l'Education Nationale et de la Recherche}.  SF and CT acknowledge
partial support from the European Community through the Research
Training Network ``The Origin of Planetary Systems'' under contract
number HPRN-CT-2002-00308, from the {\em Programme National de
Plan\'etologie} and from the {\em Programme National de Physique
Stellaire}. The simulations presented in this paper were performed at
the 
Institut du D\'eveloppement et des Resources en Informatique
Scientifique.

\bibliographystyle{apj}
\bibliography{author}

\end{document}